\documentclass[aps,superscriptaddress,nofootinbib,tightenlines,prd]{revtex4}
\usepackage[T1]{fontenc}
\usepackage[latin9]{inputenc}
\usepackage{float}
\usepackage{amsmath}
\usepackage{mathtools}
\usepackage{esint}
\usepackage{ulem}
\usepackage{amsfonts}
\usepackage{multirow}
\usepackage{mathrsfs}
\usepackage{subfig}
\usepackage{amsmath}
\usepackage{amssymb}
\usepackage{bm}
\usepackage{braket}
\usepackage[dvipsnames]{xcolor}
\usepackage{slashed}
\usepackage{graphicx}
\usepackage{caption}
\usepackage{mathbbol}
\usepackage{bbold}

\usepackage[absolute]{textpos}
\newcommand {\bseq}{\begin{subequations}}
\newcommand {\eseq}{\end{subequations}}

\newcommand*{\rom}[1]{\uppercase\expandafter{\romannumeral #1\relax}}
\def\Xint#1{\mathchoice
{\XXint\displaystyle\textstyle{#1}}%
{\XXint\textstyle\scriptstyle{#1}}%
{\XXint\scriptstyle\scriptscriptstyle{#1}}%
{\XXint\scriptscriptstyle\scriptscriptstyle{#1}}%
\!\int}
\def\XXint#1#2#3{{\setbox0=\hbox{$#1{#2#3}{\int}$ }
\vcenter{\hbox{$#2#3$ }}\kern-.6\wd0}}

\def\dashint{\Xint-}
\def\Hlfz{\mathbb{H}_{\mathrm{LF}}^{(0)}}
\def\Hz{\mathbb{H}^{(0)}}
\def\Vlf{\mathbb{V}_\mathrm{LF}}
\def\V{\mathbb{V}}

\def\QCDtwo{$\mathrm{QCD}_2$}
\def\qqquad{\quad\quad\quad}
\def\OO{\mathbb{O}}
\newcommand{\nn}{\nonumber}

\newcommand{\beq}{\begin{equation}}
\newcommand{\eeq}{\end{equation}}
\newcommand{\bqa}{\begin{eqnarray}}
\newcommand{\eqa}{\end{eqnarray}}
\newcommand{\bsq}{\begin{subequations}}
\newcommand{\esq}{\end{subequations}}

\newcommand{\QCDtw}{${\mathrm{QCD}_2}$}

\DeclareSymbolFontAlphabet{\mathbbm}{bbold}
\DeclareSymbolFontAlphabet{\mathbb}{AMSb}%

\begin{document}

\begin{textblock}{6}(15.5,0.75)
  \noindent \it{To appear in Phys. Rev. D}
  \end{textblock}

\title{Light-cone and quasi generalized parton distributions\\ in the 't Hooft model}

\author{Yu Jia\footnote{jiay@ihep.ac.cn}}
\affiliation{Institute of High Energy Physics,  Chinese Academy of Sciences, Beijing 100049, China\vspace{0.2cm}}
\affiliation{School of Physics, University of Chinese Academy of Sciences,
Beijing 100049, China\vspace{0.2cm}}

\author{Zhewen Mo~\footnote{mozw@itp.ac.cn}}
\affiliation{Institute of High Energy Physics, Chinese Academy of Sciences, Beijing 100049, China\vspace{0.2cm}}
\affiliation{CAS Key Laboratory of Theoretical Physics, Institute of Theoretical Physics,Chinese Academy of Sciences, Beijing 100190, China\vspace{0.2cm}}

\author{Xiaonu Xiong\footnote{xnxiong@csu.edu.cn}}
\affiliation{School of Physics, Central South University, Changsha 410083, China\vspace{0.2cm}}

\author{Rui Yu\footnote{yurui@csrc.ac.cn}}
\affiliation{School of Physical Science and Technology,
Inner Mongolia University, Hohhot 010021, China\vspace{0.2cm}}
\affiliation{Beijing computational Science Research Center, Beijing 100049, China\vspace{0.2cm}}


\begin{abstract}
We present a comprehensive study of the light-cone generalized parton distribution (GPD) and quasi-GPD of a flavor-neutral meson
in the 't Hooft model, {\it i.e.}, two-dimensional QCD (\QCDtw) in the $N_c\to\infty$ limit.
With the aid of the Hamiltonian approach, we construct the light-cone GPD in terms of the meson's light-cone wave function
in the framework of light-front quantization, and express the quasi-GPD in terms of the meson's Bars-Green wave functions and 
the chiral angle in the framework of equal-time quantization.
We show that, both analytically and numerically, the quasi-GPD does approach the light-cone GPD when the meson is boosted to the infinite momentum frame,
which justifies the tenet underlying the large momentum effective theory for the off-forward parton distribution.
Upon taking the forward limit, the light-cone and quasi-GPDs reduce to the light-cone and quasi-PDFs.
As a bonus, we take this chance to correct the incomplete expression of the quasi-PDFs in the 't Hooft model
reported in our preceding work [Y. Jia et al. Phys. Rev. D 98, 054011 (2018)].
\end{abstract}

\maketitle

\section{Introduction}\label{sec:intro}

Unraveling the internal structure of nucleon is one of the most important topic in QCD, and also constitutes
the major scientific goal of the prospective electron ion collision experiments,
such as electron-ion collider (EIC) and  electron-ion collider in China (EicC)~\cite{Accardi:2012qut,Anderle:2021wcy}.
The internal partonic structure of a nucleon is fruitfully characterized by various light-cone partonic distribution functions,
such as parton distribution functions (PDFs), transverse momentum dependent parton distributions and generalized parton distribution functions (GPDs).
These distribution functions are nonperturbative yet universal objects,
which serve as essential inputs for making accurate predictions for high-energy collision
experiments as required by the QCD factorization theorem.

Among these light-cone distribution functions, the GPD encodes much richer information about the partonic structure of nucleon
than the ordinary PDFs. The GPDs embody the correlation between a parton's longitudinal momentum fraction and its transverse position and thus may provide a three-dimensional tomographic portrait of a nucleon.

The GPDs can in principle be extracted through exclusive lepton-hadron scattering processes such as deeply virtual Compton scattering~\cite{Ji:1996nm}
and vector meson production processes~\cite{Collins:1996fb, Goloskokov:2005sd}.
The measurements of GPDs are among the top priority list in the projected EIC and EicC programs~\cite{Accardi:2012qut,Anderle:2021wcy}.
On the theoretical ground, there exist some estimation of the GPDs from QCD-inspired models, such as the light-cone quark model~\cite{Scopetta:2002xq}.
In contrast to the phenomenological models, lattice QCD is currently viewed as the only reliable and model-independent approach to compute GPD.
It is the Mellin moments of the GPD which can be directly handled in lattice QCD, which correspond to the off-forward
matrix elements of some local operators. In principle, the $x$ dependence of the GPD can be reconstructed once infinite towers of the Mellins moments are known.
Unfortunately, calculation of higher Mellin moments in lattice suffers from a severe operator mixing obstacle. Moreover,
high-order derivatives in the operators affiliated with the high-order Mellin moments
demands finer lattice spacing, which is also computationally expensive.
After decades-long efforts, only a few lower-order Mellin moments of the GPD have been computed on the lattice.
To the best of our knowledge, so far only the first four Mellin moments of GPD have been available from lattice simulation~\cite{Hagler:2003jd,Gockeler:2003jfa,QCDSF:2006tkx,LHPC:2007blg,Hannaford-Gunn:2022lez}.

A theoretical breakthrough in the last decade is the advent of Large Momentum Effective field Theory (LaMET), which enables one to directly extract
the $x$ dependence of the light-cone parton distributions on the lattice~\cite{Ji:2013dva,Ji:2020ect}.
In the LaMET framework, a central object is called quasi-GPD, which is defined as the equal-time yet spatially nonlocal operator matrix element
where the external nucleon states carry finite momenta, thus can be directly accessed by lattice simulation.
The $x$ dependence of the light-cone GPD can then be inferred from the quasi-GPD through the
perturbative matching procedure. Very recently, there have emerged some exploratory lattice studies of the GPD
following the LaMET approach. While most of the investigation concentrate on the zero-skewness ($\xi=0$) case~\cite{Bhattacharya:2022aob,Chen:2019lcm,Lin:2021brq},
some study also considers the nonzero skewness case ($|\xi|=1/3$)~\cite{Alexandrou:2020zbe}.

Because of greater complexity, there is still a long path to go to reconstruct the whole profiles of the quasi-GPDs and light-cone GPDs,
even in the framework of LaMET.
In the meanwhile, it may look appealing if one can learn some lessons from toy models of QCD in which the quasi-GPDs and GPDs may be rigorously computed.
In fact, the 't Hooft model, {\it i.e.}, the two-dimensional QCD $N_c\to\infty$ limit~\cite{tHooft:1974pnl},
is an ideal theoretical laboratory to investigate the light-cone and quasi
parton distributions of a meson. As a solvable model, {\QCDtw} resembles the realistic QCD in several aspects,
such as color confinement, Regge trajectories, ``naive'' asymptotic freedom, nonzero quark condensate, {\it etc.}.
Recently the quasi-PDFs~\cite{Jia:2018qee}\cite{Ji:2020ect} and intrinsic charm PDF of light mesons~\cite{Hu:2022wsf} have been investigated in this model.
GPDs are considerably more complicated than the PDFs, since it depends on three kinematic variables instead of a single momentum fraction variable.
The light-cone GPD of a charged meson has been investigated in the 't Hooft model by Burkardt in 2000~\cite{Burkardt:2000uu}.
It is the goal of this work to carry out a comprehensive investigation on
both the light-cone and quasi-GPDs of a flavor-neutral meson with various quark mass.
Employing the Hamiltonian approach, we are able to express the light-cone GPD in terms of the meson's light-cone wave function (LCWF)
through light-front quantization, and express the quasi-GPD in terms of the meson's Bars-Green wave functions (BGWFs) through equal-time quantization.
The highlight of this work is that, when the meson is boosted faster and faster, the quasi-GPD does converge to the light-cone GPD.
We exhibit this feature in both analytical and numerical manner. Therefore, our study corroborates the
key assumption of LaMET in {\QCDtwo}, that the quasi partonic distributions smoothly transition into the light-cone counterparts
when a hadron is boosted from the finite momentum to the infinite momentum frame (IMF).

The rest of the paper is organized as follows.
In Sec.~\ref{sec:gpd_rev}, we recap the definitions and some basic properties of the light-cone GPDs and
quasi-GPDs, and explain how these definitions are adapted to two-dimensional spacetime.
In Sec.~\ref{Sec:review:tHooftmodel:Hamiltonian:approach}, we review the Hamiltonian approach in the 't Hooft model.
Concretely speaking, we recap the bosonization procedures in the light-front quantization as well as in equal-time quantization,
and how to arrive at the respective bound-state equations: the 't Hooft equation and Bars-Green equations. We also organize the Hamiltonian
in $1/N_c$ expansion, including the ${\cal O}(1/\sqrt{N_c})$ terms in both light-front and equal-time quantization.
In Secs.~\ref{Sec:LCGPD:derivation} and \ref{Sec:Quasi:GPD:derivation}, we employ the Hamiltonian approach
to construct the explicit expressions of the light-cone and quasi-GPDs of a flavor-neutral meson,
at the lowest order in $1/N_c$.
In Sec.~\ref{Sec:Infinite:momentum:limit} we analytically prove that the light-cone GPD can be reached from the
quasi-GPD in the infinite momentum limit.
In Sec.~\ref{sec:foward_limt}, as a byproduct, we obtain the expressions of the quasi-PDFs
by taking the the forward limit $\varDelta\to 0$ of the quasi-GPDs.
We devote Sec.~\ref{sec:num} to a comprehensive numerical study of the light-cone and quasi-GPDs, with different choices of quark mass and skewness.
We also show the profiles of various quasi-PDFs with different quark masses.
Finally we summarize in Sec.~\ref{sec:sum}.
In Appendices~\ref{appendix:A} and~\ref{appendix:B},
we present some lengthy formulas for the interacting Hamiltonian and three-meson vertex functions in light-front and equal-time quantization, respectively.
In Appendix~\ref{appendix:C}, we make a comparative numerical study for the quasi-PDFs of flavor-neutral mesons
between our new correct results and the incomplete old results.

\section{Brief Review of light-cone and quasi-GPDs}
\label{sec:gpd_rev}

In this section, we recap the definitions and some key properties of the light-cone GPD and quasi-GPD.
For more comprehensive reviews on light-cone GPD, we refer the interested readers to Refs.~\cite{Ji:1998pc,Diehl:2003ny}.
GPDs generalize PDFs to the nonforward kinematics where the momentum carried by the final-state hadron differs from that carried by the initial-state hadron.

The unpolarized quark GPD inside a spin-${1\over 2}$ nucleon is defined as
\begin{align}
F_q(x,\xi,t) =\!\! \int\frac{d\eta^-}{4\pi} e^{i xP^+\eta^-}\!\!\left\langle P\!+\!\frac{\varDelta}{2}\right|&\overline{\psi}\left(\!-\frac{\eta^-}{2}\!\right)\!\mathcal{P}\left\{\!\exp\!\left[-ig_s\int_{-\frac{\eta^-}{2}}^{\frac{\eta^-}{2}}\!\! d\zeta^-\! A^{+,a}(\zeta^-)\mathrm{t}^a\right]\!\right\}
\gamma^+{\psi}\left(\!\frac{\eta^-}{2}\!\right)\left|P\!-\!\frac{\varDelta}{2}\right\rangle  \label{eq:lcgpd_nucl_def}
\end{align}
where $\psi$ denotes the quark field, $\mathrm{t}^a$ signifies the SU(3) generator in fundamental representation, and the path-ordered
exponential $\mathcal{P}\left\{\exp[\cdots]\right\}$ represents the gauge link which ensures the gauge invariance of the GPD.
$x$ denotes the light-cone momentum fraction,  $\xi = \varDelta^+/(2P^+)$ is known as the skewness parameter, and
 $t=\varDelta^2$ is the square of hadron momentum transfer~\footnote{
 We emphasize here we follow the convention of \cite{Ji:2015qla,Xiong:2015nua} to define $\xi$, which differs from some literature by an extra minus sign.
 The different convention in defining $\xi$ does not affect the results of this work, because the GPDs {discussed in this work} are even functions of $\xi$
 due to the time-reversal invariance~\cite{Ji:1998pc,Diehl:2003ny,Belitsky:2005qn}.}.

 The quark GPD of a nucleon can be decomposed into two pieces:
 \begin{align}
 F_q(x,\xi,t) = H_q(x,\xi,t)\overline{u}\left(P+\frac{\varDelta}{2}\right)\gamma^+u\left(P-\frac{\varDelta}{2}\right) + E_q(x,\xi,t)\overline{u}\left(P+\frac{\varDelta}{2}\right)\frac{\sigma^{+\rho}\varDelta_\rho}{2M_N} u\left(P-\frac{\varDelta}{2}\right),
 \label{eq:H_E_type}
 \end{align}
 where $u(P\pm\frac{\varDelta}{2})$ denotes the incoming and outgoing nucleon Dirac spinors, and
 the $H$-type and $E$-type GPDs are related to the nucleon electromagnetic form factor $F_{1,2}$ via~\cite{Ji:1996ek}
 \begin{align}
 \sum_q \int dx H_q(x,\xi,t) = F_1(t),\quad \sum_q \int dx E_q(x,\xi,t) = F_2(t).
 \end{align}

GPDs entail much richer information about how partons are distributed inside a nucleon than PDF, since they depend on
on two extra kinematical variables, $\xi$ and $t$. 
In the special case of $\xi=0$,  if only the transverse components of $\varDelta$ are nonzero,
 the zero-skewness GPD is related to the combined distributions of
 parton's light-cone momentum fraction $x$ and transverse position $\boldsymbol{b}_\perp$ (impact parameter space) through a two-dimensional Fourier transformation
 with respect to $\varDelta_\perp$~\cite{Burkardt:2002hr},
 \bsq
 \begin{align}
 \mathcal{H}_q(x,\bm b_\perp)=&\int \frac{d\bm\varDelta_\perp^2}{(2\pi)^2}e^{-i\boldsymbol{\varDelta}_\perp\cdot\boldsymbol{b}_\perp}H_q(x,\xi\!=\!0,t\!=\!-\boldsymbol{\varDelta}_\perp^2),\\ \mathcal{E}_q(x,\bm b_\perp)=&\int \frac{d\bm\varDelta_\perp^2}{(2\pi)^2}e^{-i\boldsymbol{\varDelta}_\perp\cdot\boldsymbol{b}_\perp}E_q(x,\xi\!=\!0,t\!=\!-\boldsymbol{\varDelta}_\perp^2),
 \end{align}
 \esq
where $\mathcal{H}_q(x,\bm b_\perp)$ can be interpreted as the probability density of finding a parton carrying light-cone momentum fraction $x$ at transverse position $\bm b_\perp$ and
$\mathcal{E}_q(x,\bm b_\perp)$ characterizes the distortion of the parton distribution in $x,\boldsymbol{b}_\perp$ space induced by nucleon's spin effects~\cite{Burkardt:2002hr}.

 GPD also entails information about the parton's angular momentum distribution.
 For instance, the celebrated Ji's sum rule reveals the profound connection between GPDs and nucleon spin~\cite{Ji:1996ek}:
 \beq
 S=\sum_{f=q,g} J_f=\sum_{f=q,g}\frac{1}{2}\int dx \,x \,\left[H_f(x,0,0) +E_f(x,0,0)\right],~\label{eq:sumrule}
 \eeq
 where $J_f$ denotes the quark and gluon contributions to the nucleon spin.

In the forward limit $\xi\rightarrow 0, t\rightarrow 0$, the $H$-type GPD of the nucleon in \eqref{eq:H_E_type} reduces to the collinear PDF, and the
$E$-type GPD is related to the angular momentum of the parton, as indicated in the spin sum rule~\eqref{eq:sumrule}.

In the nonvanishing skewness case, the $H$-type quark GPD satisfies the following positivity bound in the DGLAP region ($\xi<x<1$) ~\cite{Pire:1998nw}
 \beq
 |H_q(x,\xi,t)|\leq \sqrt{q\left(\frac{x-\xi}{1-\xi}\right)q\left(\frac{x+\xi}{1+\xi}\right)},
 \label{eq:positivity}
 \eeq
 where $q(x)$ signifies the quark PDF. The derivation of the positivity bound involves the overlap representation of GPD~\cite{Radyushkin:1998es,Diehl:2003ny,Belitsky:2005qn}.

 Another peculiar trait of GPD is the polynomiality, which is a direct consequence of Lorentz symmetry. Polynomiality of GPD states that
 the GPD's $n$th Mellin moment is an $n$th-order polynomial of $\xi$ (for a detailed discussion, see \cite{Ji:1998pc,Diehl:2003ny,Belitsky:2005qn}).

For a spinless meson exemplified by $\pi$, only the $H$-type GPD survives~\cite{Meissner:2008ay}:
 \begin{align}
   H_q(x,\xi,t) =\!\! \int\frac{d\eta^-}{4\pi} e^{i xP^+\eta^-}\!\!\left\langle P\!+\!\frac{\varDelta}{2}\right|\bar{\psi}\left(\!-\frac{\eta^-}{2}\!\right)\!\mathcal{P}\left\{\!\exp\!\left[-ig_s\int_{-\frac{\eta^-}{2}}^{\frac{\eta^-}{2}}\!\! d\zeta^-\! A^{+,a}(\zeta^-)\mathrm{t}^a\right]\!\right\}
   \gamma^+{\psi}\left(\!\frac{\eta^-}{2}\!\right)\left|P\!-\!\frac{\varDelta}{2}\right\rangle.
\label{eq:lcgpd_s0}
 \end{align}
 As we will explain, this definition is GPD is most relevant for our study in the meson's GPD in the 't Hooft model.

 The quark quasi-GPD of a spin-0 meson is defined as the purely spatial correlator~\cite{Ji:2013dva,Ji:2020ect}:
 \begin{align}
   \tilde{H}_q(x,\xi,t,P_z) = \int\frac{dz}{4\pi} e^{-i xP^zz}\left\langle P\!+\!\frac{\varDelta}{2}\right|\bar{\psi}\left(-\frac{z}{2}\right)\mathcal{P}\left\{\!\exp\!\left[-ig_s\int_{-\frac{z}{2}}^{\frac{z}{2}}\!\! d\zeta\! A^{z,a}(\zeta)\mathrm{t}^a\right]\right\}\gamma^z{\psi}\left(\frac{z}{2}\right)\left|P\!-\!\frac{\varDelta}{2}\right\rangle,
 \label{eq:quasigpd_s0}
 \end{align}
where $x=k^z/P^z$ signifies the ratio of the longitudinal momentum of the quark to the average longitudinal momentum between the initial and final-state meson,
$\xi = \varDelta^z/(2P^z)$ denotes the skewness parameter in quasi-GPD~\cite{Ji:2015qla,Xiong:2015nua}, and $t=\varDelta^2$.
The path-ordered exponential denotes the gauge link along the $z$ direction.

The key idea of the LaMeT is that the light-cone GPD and quasi-GPD share  identical infrared behavior yet differ in the ultraviolet.
The difference can be compensated by a perturbative matching factor. For instance, the $H$-type quasi-GPD is linked with the  $H$-type light-cone GPD
via the following factorization theorem~\cite{Ji:2015qla,Xiong:2015nua}:
 \begin{align}
 \widetilde{H}(x,\xi,t,P_z) =&\int_{-1}^{1}\frac{dy}{y} Z\left(\frac{x}{y},\frac{\xi}{y},t\right) H(y,\xi,t)+\cdots,
 \label{eq:gpd_matching}
 \end{align}
 where $Z$ signifies the perturbative calculable  short-distance coefficient function, and the ellipses represents the higher-twist correction
 suppressed by powers of $1/P_z$. It is the quasi-GPD that can be computed on the lattice, and subsequently one can extract the light-cone GPD
 by inverting the matching formula \eqref{eq:gpd_matching}.

In the rest of the work, we will consider the quark light-cone GPD and quasi-GPD of a meson in the 't Hooft model
(Since we are dealing with the $N_c\to\infty $ limit, we refrain from considering the GPD of an infinitely-heavy baryon).
In 1+1-dimensional spacetime, there is no such notion as  angular momentum (orbital or spin); therefore, the 3+1-dimensional definitions of quark GPD of a spin-0 meson,
\eqref{eq:lcgpd_s0} and the quark quasi-GPD of a spin-0 meson \eqref{eq:quasigpd_s0} can be directly carried over to {\QCDtwo}.
Moreover, the absence of transverse degree of freedom brings in additional simplification-that the skewness parameter $\xi$ and the squared momentum transfer $t$ actually are related to each other:
\bseq
\begin{itemize}
 \item light-cone GPD
 \beq
 t \equiv \varDelta^2=\frac{4\mu_n^2\xi^2}{\xi^2-1},\label{eq:t_xi_lc}
 \eeq
 \item quasi-GPD
\beq
 t \equiv \varDelta^2 = 2 \left(\mu_n^2+(1-\xi^2)P_z^2-\sqrt{\left(\mu_n^2+P_z^2 \left(\xi ^2+1\right)\right)^2-4\xi ^2P_z^4 }\right).\label{eq:t_xi_quasi}
\eeq
\end{itemize}
 \eseq
where $\mu_n$ signifies the mass of the $n$th excited mesonic state in the 't Hooft model. The incoming and outgoing meson states satisfy the on-shell condition
$(P\mp\frac{\varDelta}{2})^2=\mu_n^2$.
It is straightforward to verify that \eqref{eq:t_xi_quasi} reduces to \eqref{eq:t_xi_lc} in $P_z\rightarrow \infty$ limit.

Because of the absence of transverse spatial degree in {\QCDtwo}, after imposing the physical gauge such as light-cone gauge $A^{+,a}=0$,
the $A^{-,a}$ field becomes a constrained, rather than dynamical variable.
Hence the role played by the gluon is to provide an
instantaneous interquark linear Coulomb potential.
As a consequence, the light-cone gluon GPD trivially vanishes. Therefore, in this work, we
concentrate on the quark sector of GPDs. Without causing confusion, we will
simply drop the subscript ``$q$'' in the quark light-cone and quasi-GPDs henceforth.

\section{A brief review of hamiltonian approach of 't Hooft model}\label{sec:bnd_state_eq}
\label{Sec:review:tHooftmodel:Hamiltonian:approach}

Our starting point is the {\QCDtwo} Lagrangian with a single quark flavor:
 \beq
     \mathcal{L} =
     - \frac{1}{4}F^{\mu\nu,a}F^{a}_{\mu\nu}
     + \overline\psi\left(i\slashed{D} - m\right)\psi,
    \label{QCD:lagr}
 \eeq
with $D_\mu= \partial_\mu-ig_s A_\mu^a\mathrm{t}^a$ signifying color covariant derivative,  and $\mathrm{t}^a$ denoting the generators of the $\mathrm{SU}(N_c)$ group in the fundamental representation.
 The gluon field strength tensor is defined as $F_{\mu\nu}^a \equiv \partial_\mu A_\nu^a-\partial_\nu A_\mu^a+g_sf^{abc}A_\mu^bA_\nu^c$.
 We adopt the chiral-Weyl representation for the Dirac-$\gamma$ matrices:
 \begin{align}
 \gamma^0 = \sigma_1,\qquad \gamma^z=-i\sigma_2,\qquad \gamma_5=\gamma^0\gamma^z=\sigma_3,
 \end{align}
 where $\sigma_i$ ($i=1,2,3$) are Pauli matrices.

To make nonperturbative dynamics more tractable, we also resort to $1/N_c$ expansion. It is convenient to introduce the 't Hooft coupling constant
$\lambda={g_s^2N_c}/{4\pi}$,  which bears mass dimension 2. We are interested in the limit where $N_c\rightarrow\infty$ but $\lambda$ is kept fixed.

The 't Hooft model can be solved in both diagrammatic and Hamiltonian approaches.  The bound-state equation arising from the
light-front quantization is dubbed 't Hooft equation, which describes the meson viewed in the IMF.
In contrast, the bound-state equations arising from the equal-time quantization are called Bars-Green equations,
which characterize mesons viewed in a finite momentum frame (FMF).
In the rest of this section, we will present a brief review about the Hamiltonian approach in both light-front and equal-time quantization,
which constitutes the essential prerequisites to derive the functional forms of the light-cone GPD and quasi-GPD.
See Refs.~\cite{Kikkawa:1980dc,Nakamura:1981zi,Itakura:1996bk,Bars:1977ud,Jia:2017uul,Jia:2018qee} for detailed introduction of Hamiltonian approach.

\subsection{Hamiltonian approach in light-front quantization}
\label{H:approach:LF:quantization}

We introduce the light-cone coordinates as $\xi^{\pm} = ({\xi^0\pm\xi^z})/{\sqrt{2}}$ and express the Dirac spinor field as
\begin{equation}
    \psi = 2^{-{1\over 4}}
    \left(
    \begin{array}{c}
        \psi_R \\ \psi_L
    \end{array}
    \right),
    \label{psidcp}
\end{equation}
where $R$, $L$ denote the right-handed and left-handed components.

Substituting \eqref{psidcp} into \eqref{QCD:lagr}, and imposing the light-cone gauge $A^{+, a} = 0$,
one can express $A^{-,a}$ and $\psi_{L}$ as functions of  $\psi_R$ from equation of motion~\footnote{In {\QCDtw}, the right-hand spinor is equivalent to the ``good'' component, because the projection operator $\frac{1}{2}\gamma^-\gamma^+$ coincides with $\frac{1+\gamma_5}{2}$.}~\cite{tHooft:1974pnl}.
After Legendre transformation, one arrives at the light-front (LF) Hamiltonian solely in terms of $\psi_R$~\cite{tHooft:1974pnl}.
The theory can then be canonically quantized in equal light-front time. The canonical quantization rules in equal light-front time are then
\bseq
\bqa
&& \left \{\psi^i_R(x^-),{\psi^j_R}^\dagger(y^-) \right\} =\delta^{ij}\delta (x^--y^- ),
\\
&& \left  \{\psi^i_R(x^-),\psi^j_R(y^-) \right \} = \left\{ {\psi^i_R}^\dagger(x^-),
{\psi^j_R}^\dagger(y^-)\right\} =0,
\eqa
\eseq
with $i,j=1,\ldots, N_c$ indicating the color indices.

At $x^+=0$, the right-handed quark field can be Fourier expanded as follows:
\beq
\psi_R^{i}(x^-)=\int_0^\infty\frac{dk^+}{2\pi}\left[b^i(k^+)e^{-ik^+x^-}+d^{i\dagger}(k^+)e^{ik^+x^-}\right].
\label{psiR:Four:expand}
\eeq
The quark(antiquark) annihilation/creation operator $b/b^\dagger$ ($d/d^\dagger$)
obeys the standard anticommutation relations:
\begin{equation}
\{ b^{i\dagger}(k^+), b^j(p^+)\} = 2\pi \delta^{ij} \delta(k^+-p^+),  \qquad  {\{d^{i \dagger}(k^+), d^j(p^+)\} = 2\pi \delta^{ij} \delta(k^+-p^+)},
\label{eq:bb}
\end{equation}
and all other unspecified anticommutators simply vanish.

Substituting \eqref{psiR:Four:expand} into the light-front Hamiltonian, one encounters various
bilinear terms composed of quark/antiquark annihilation and
creation operators.  It is convenient to adopt the bosonization technique to facilitate the diagonalization of Hamiltonian~\cite{Kikkawa:1980dc,Nakamura:1981zi,Rajeev:1994tr,Dhar:1994ib,Dhar:1994aw,Cavicchi:1993jh,Barbon:1994au,Itakura:1996bk}
by introducing the following bosonic compound operators~\footnote{Note the normalization of the compound operators $B$ and $D$ here
differs from what is given in our previous work~\cite{Jia:2018mqi}. The purpose of making this change is to
make the $1/N_c$ expansion of the Hamiltonian manifest.}:
\bseq
\begin{align}
M(k^+,p^+)&=\frac{1}{\sqrt{N_c}}\sum_i d^i(k^+)b^i(p^+),\\
B(k^+,p^+) &= \sum_i b^{i\dagger}({k^+})b^i({p^+}), \\
D(k^+,p^+) &= \sum_i d^{i\dagger}({k^+})d^i({p^+}).
\end{align}
\label{Def:LF:bosonic:compound}
\eseq

In the following, the LF Hamiltonian of the 't Hooft model will be reexpressed in terms of the bosonized operators such as mesonic annihilation and creation operators.
The LF Hamiltonian can be organized in powers of $1/N_c$:
\begin{align}
  \mathbb{H}_{\mathrm{LF}} = \mathbb{H}_{\mathrm{LF,vac}}+ \mathbb{H}_{\mathrm{LF}}^{(0)} + \mathbb{V}_{\mathrm{LF}}.
\label{eq:HLF_full}
\end{align}

The leading $\mathcal{O}(N_c)$ piece, $\mathbb{H}_{\mathrm{LF}}^{(0)}$, corresponds to the vacuum energy, which is badly UV and IR divergent~\cite{Lenz:1991sa}.
However, since it is proportional to the unit operator and does not have any physical impact,
we will simply discard $\mathrm{H_{LF,vac}}$ henceforth.

The $\mathcal{O}(N_c^0)$ piece in \eqref{eq:HLF_full} involves the integration over $B+D$ and $M^\dagger M$.
A key insight is to realize that those bosonic compound operators in \eqref{Def:LF:bosonic:compound}
are not independent.
In fact $B$ and $D$ can be expressed as the convolution between $M$ and $M^\dagger$:
\bseq
\begin{align}
B(k^+,p^+)&=\int_0^\infty \frac{dq^+}{2\pi} M^{\dagger}(q^+,k^+)M(q^+,p^+),
\\
D(k^+,p^+)&=\int_0^\infty \frac{dq^+}{2\pi} M^{\dagger}(k^+,q^+)M(p^+,q^+).
\end{align}
\label{eq:BD2M}
\eseq
The reason is that, in a confining theory like \QCDtwo,
one cannot create an isolated quark or antiquark from the vacuum and rather only creates a color-singlet quark-antiquark pair from the vacuum~\cite{Kalashnikova:2001df}.
Substituting \eqref{eq:BD2M} into \eqref{eq:HLF_full}, the LF Hamiltonian can be expressed solely in terms of $M$ and $M^\dagger$.
$M$ and $M^\dagger$ satisfy the simple commutation relations:
\beq
 \left[M(k_1,p_1),M^\dagger(k_2,p_2)\right] = (2\pi)^2\delta(k_1-k_2)\delta(p_1-p_2) + {\cal O}\left(\frac{1}{N_c}\right).
 \label{eq:MM_com}
 \eeq

To facilitate  the diagonalization of $\mathbb{H}_{\mathrm{LF}}^{(0)}$,  it is convenient to introduce a new set of mesonic annihilation/creation operators $m_n/m_n^\dagger$,
which are related to $M$ and $M^\dagger$ through
\bseq
\begin{align}
&m_n(P^+)=\sqrt{\frac{P^+}{2\pi}}\int_0^1 dx\phi_n(x)M((1-x)P^+,xP^+),
\\
&M((1-x)P^+,xP^+)=\sqrt{\frac{2\pi}{P^+}}\sum_{n=0}^{\infty}\phi_n(x)m_n(P^+),
\end{align}
\label{eq:m2M}
\eseq
where the coefficient functions $\phi_n(x)$ later will be identified with the 't Hooft LCWF of the $n$th excited mesonic state, with $x\in (0,1)$ denoting the light-cone momentum fraction carried by the quark inside the meson.

If the mesonic annihilation and creation operators are required to obey the standard commutation relation:
\beq
\left[m_n(P_1^+),m^\dagger_r(P_2^+)\right]=2\pi\delta_{nr}\delta(P_1^+-P_2^+),
\label{LF:mesonic:operators:commutator}
\eeq
the 't Hooft wave functions must obey the following orthogonality and completeness conditions:
\bseq
\begin{align}
&\int_0^1 dx \phi_n(x)\phi_r(x)=\delta_{nr},
\\
&\sum_n \phi_n(x)\phi_n(y)=\delta(x-y).
\end{align}
\eseq

Substituting \eqref{eq:BD2M}, \eqref{eq:m2M} into \eqref{eq:HLF_full},
our goal is to put the $\mathbb{H}_{\mathrm{LF}}^{(0)}$ into a diagonalized form, which describes
an infinite towers of noninteracting mesons:
\beq
\mathbb{H}_{\mathrm{LF}}^{(0)}= \sum_n \int_0^\infty\frac{dP^+}{2\pi}P_n^-m_n^\dagger(P^+)m_n(P^+),
\eeq
where $P_n^-=M_n^2/(2P^+)$ denotes the light-cone energy of the $n$th excited state meson with light-cone momentum $P^+$.

To fulfill this goal, one must enforce that all the off-diagonal terms in $\mathbb{H}_{\mathrm{LF}}^{(0)}$ cancel,
which in turn imposes the following constraints on infinite numbers of coefficient functions $\phi_n(x)$:
\beq
\left(\frac{m^2\!-\!2\lambda}{x}\!+\!\frac{m^2\!-\!2\lambda}{1-x}\!-\!M_n^2\right)\phi_n(x) = 2\lambda
\dashint_0^1 \frac{dy}{(x-y)^2}\phi_n(y).
\label{eq:tHooft_eq}
\eeq
This is nothing but the celebrated 't Hooft equation in \QCDtwo, the bound-state equation for the $n$th excited mesonic state
in the 't Hooft model in IMF.
Note that the dashed integral $\int\!\!\!\!\!-$ in \eqref{eq:tHooft_eq} signifies the principle-value prescription,
whose role is to tame the IR divergence as $y\to x$.

The single mesonic state can be constructed as
\begin{align}
\left|P^+ \right\rangle = \sqrt{2P^+}m_n^\dagger(P^+)\left|0\right\rangle.
\label{single:meson:LF:quantization}
\end{align}

The last operator $\mathrm{V}$ in the LF Hamiltonian in \eqref{eq:HLF_full} scales as ${\cal O}(1/\sqrt{N_c})$,  which involves integration
of the triple product of $m$ and $m^\dagger$.
For the purpose of computing the quark light-cone GPD, it is sufficient to know
 \begin{align}
\notag \Vlf=&\frac{\lambda}{(2\pi)^{3/2}\sqrt{N_c}}\sum_{n_1,n_2,n_3}\int_0^{\infty} dk_1^+dk_2^+dk_3^+dk_4^+dq_1^+\Bigg[\frac{-\delta \left(k_1^++k_2^+-k_3^++k_4^+\right) }{\left(k_2^++k_4^+\right)^2\sqrt{{\left(k_1^++k_2^+\right) \left(k_3^++q_1^+\right) \left(k_4^++q_1^+\right)}}}
 \\
      &\times  \varphi_{n_1}\left(\frac{k_1^+}{k_1^++k_2^+}\right) \varphi_{n_2}\left(\frac{k_3^+}{k_3^++q_1^+}\right) \varphi_{n_3}\left(\frac{k_4^+}{k_4^++q_1^+}\right) m_{n_1}\left(k_1^++k_2^+\right)m_{n_2}^{\dagger }\left(k_3^++q_1^+\right)m_{n_3}\left(k_4^++q_1^+\right) +\cdots \Bigg].
\label{V:LF:one:typical:example}
\end{align}
The complete expression of $\Vlf$ is given in Appendix.~\ref{appendix:A}.
These operators induce one meson to transition into two mesons, or vice versa,
which represents an ${\cal O}(1/\sqrt{N_c})${~\cite{Callan:1975ps} effect.

\subsection{Hamiltonian approach in equal-time quantization}
\label{H:approach:ET:quantization}

To describe a moving meson with a finite momentum, it is more appropriate to adopt the equal-time quantization rather than light-front quantization.
In 1978 Bars and Green solved the 't Hooft model from this perspective~\cite{Bars:1977ud}. Upon imposing the axial gauge $A^{1,a}=0$,
Employing the Euler-Lagrange equation,  one can express $A^{0,a}$ as a functional of $\psi$ and $\psi^\dagger$.
After Legendre transformation, one arrives at the Hamiltonian solely in terms of $\psi$ and $\psi^\dagger$.
The canonical quantization rule at equal time reads
\bseq
\bqa
&& \left \{\psi^i(z),{\psi^j}^\dagger(z')\right \} = \delta_{ij}\delta (z-z'),
\\
&& \left \{\psi^i(z),\psi^j(z')\right \} =\left\{{\psi^i}^\dagger(z),{\psi^j}^\dagger(z')\right\}=0.
\eqa
\eseq

At $t=0$, the quark Dirac field can be Fourier expanded as follows~\footnote{In the cases where no confusion can arise,
we often frequently suppress the superscript $``z"$ of a 2-vector
to condense the notation. Therefore, in most cases, $k$ is the shorthand for $k^z$, the spatial component of a
2-momentum $k^\mu=(k^0, k^z)$.}:
\begin{align}
 \psi^i(z) = & \int_{-\infty}^\infty \frac{dk}{2\pi}\frac{1}{\sqrt{2\mathcal{E}(k)}}\left[ b^i(k) u(k)e^{ik z} + v(k)d^{i\dagger}(k)e^{-ikz} \right].
\label{Equal:time:field:expansion}
\end{align}
Here $\mathcal{E}(k)$ signifies the dressed quark energy.
The spinor wave functions in \eqref{Equal:time:field:expansion} are parametrized by
 \begin{equation}
     u(p) = \sqrt{\mathcal{E}(p)}T(p)\left( \begin{array}{c}
          1
          \\
          1
     \end{array} \right)\quad \text{ and }\quad v(p) = \sqrt{\mathcal{E}(p)}T(-p)\left( \begin{array}{c}
          1
          \\
          -1
     \end{array} \right),
 \end{equation}
 where $T(p)$ is a unitary matrix parametrized as
 \begin{align}
 T(p)=\exp\left[-\frac{1}{2}\theta(p)\gamma^z\right],
 \end{align}
with $\theta(p)$ denoting the quark chiral angle~\cite{Bars:1977ud}.
The dressed quark and antiquark annihilation operators in \eqref{Equal:time:field:expansion} annihilate the quark vacuum,
$b^i({k})\left|0\right\rangle = d^i({k})\left|0\right\rangle =0$, for all possible values of $k$.

Substituting the Fourier expansion of $\psi$ \eqref{Equal:time:field:expansion}
into the Hamiltonian, and rearranging it into the normal-ordered form,
we decompose the Hamiltonian into three pieces:
\beq
 \mathbb{H} = \mathbb{H}_0+:\mathbb{H}_2: +:\mathbb{H}_4:
\label{Hamilton:three:parts:axial}
\eeq
which contain 0, 2 and 4 quark creation/annihilation operators accordingly.

By minimizing the vacuum energy $\mathbb{H}_0$ in variation with $\theta(p)$ or, equivalently, by demanding $:\mathbb{H}_2:$ of the diagonalized form in the basis of the dressed quark and antiquark, one arrives at the so-called {\it mass-gap} equation~\cite{Bars:1977ud}
\beq
\label{formal:mass:gap:eq}
p\cos\theta(p)-m\sin\theta(p) = \frac{\lambda}{2} \dashint_{-\infty}^{\infty}\!\!
{dk\over (p-k)^2} \sin\left[\theta(p)-\theta(k)\right],
\eeq
and the dressed quark possesses the following dispersion relation:
\beq
\label{regular:dispersion:relation}
\mathcal{E}(p) \equiv   m\cos\theta(p)+p\sin\theta(p)
+\frac{\lambda}{2} \dashint_{-\infty}^{+\infty} \frac{dk}{(p-k)^2} \cos\left[\theta(p)-\theta(k)\right].
\eeq

To derive the bound state equation, we must take the $:\mathbb{H}_4:$ piece into account.
In parallel with the bosonization procedure for the LF Hamiltonian, it is useful to
introduce the following  color-singlet compound operators analogous to (\ref{Def:LF:bosonic:compound}):
\bseq
\begin{align}
M(k,p)&=\frac{1}{\sqrt{N_c}}\sum_i d^i(-k)b^i(p),\\
B(k,p) &= \sum_i b^{i\dagger}({k})b^i({p}), \\
D(k,p) &= \sum_i d^{i\dagger}({k})d^i({p}),
\end{align}
\label{Def:bosonic:compound}
\eseq

The commutation relations between $M$ and $M^\dagger$ reads
\beq
\left[M\left(k_1,p_1\right),M^\dagger\left(k_2,p_2\right)\right]=
\left(2\pi\right)^2\delta(k_1\!-\!k_2)\delta(p_1\!-\!p_2)+\,\mathcal{O}\left(\frac{1}{N_c}\right),
\label{eq:MBD:commutation:equal:time}
\eeq

Because of the confinement nature of ${\rm QCD}_2$, the same consideration
that leads to \eqref{eq:BD2M} can also be applied here {\it i.e.}, not all compound operators in
\eqref{Def:bosonic:compound} are independent.
In fact, one finds that~\cite{Kalashnikova:2001df}
\bseq
\begin{align}
B(p,p')=& \int_{-\infty}^{+\infty} \frac{dq}{2\pi} M^\dagger(q,p)M(q,p'),
\\
D(p,p')=& \int_{-\infty}^{+\infty} \frac{dq}{2\pi} M^\dagger(p,q)M(p',q).
\end{align}
\label{B:D:not:independent}
\eseq

Expressing everything in \eqref{Hamilton:three:parts:axial}in terms of the bosonic compound
operators introduced in \eqref{Def:bosonic:compound}, eliminating $B$ and $D$ in line with
(\ref{B:D:not:independent}), the Hamiltonian becomes the functional of $M$ and $M^\dagger$ and the chiral angle.
In the following the Hamiltonian of the 't Hooft model will be reexpressed in terms of the bosonized operators such as mesonic annihilation and creation operators.
One can rearrange the full Hamiltonian according to the power of $1/N_c$:
\beq
\mathbb{H} = \mathbb{H}_{\mathrm{vac}}+ \mathbb{H}^{(0)} + \mathbb{V},
\label{Hamiltonian:re:split:three:pieces}
\eeq
where $\mathbb{H}_{\mathrm{vac}}$ corresponds to the shifted vacuum energy that scales as $\mathcal{O}(N_c)$. Since it has no physical effect,
we just simply drop this constant piece.  ${\mathbb H}^{(0)}$ scales as $\mathcal{O}(N_c^0)$ and $\mathbb{V}$ scales as $\mathcal{O}(1/\sqrt{N_c})$.

To put ${\mathbb H}^{(0)}$ in the diagonal form, one can borrow the Bogoliubov transformation that is used to diagonalize the
Hamiltonian of dilute weakly interacting Bose gas~\cite{Schwabl:1997gf}, by introducing a new set of
annihilation and creation operators $m_n$ and $m_n^\dagger$ ($n=0,1,\dots$) as the linear combination of the $M$ and $M^\dagger$ operators in
\eqref{Def:bosonic:compound}~\cite{Kalashnikova:2001df}:
\bseq
\label{eq:m:mdagger:M:Mdagger:equal:time}
\begin{align}
&m_n(P)=\int_{-\infty}^{+\infty}\frac{dq}{2\pi}\left[M(q-P,q)\,\varphi^+_n(q,P)+M^\dagger(q,q-P)\,\varphi^-_n(q,P)\right],
\label{mn:expressed:in:M}
\\
&M(q-P,q) = {\frac{2\pi}{|P|}}\sum_{n=0}^\infty \left[m_n(P)\varphi_n^+(q,P)-m_n^\dagger(-P) \varphi_n^-(q-P,-P)\right],
\label{M:expressed:in:mn}
\end{align}
\label{M;mn:basis:transformation}
\eseq
where $m_n(P)$ and $m_n^\dagger(P)$ will be interpreted as the annihilation and
creation operators for the $n$th mesonic state carrying spatial momentum $P$.
The functions $\varphi_n^+(q,P)$ and $\varphi_n^-(q,P)$ play the role of Bogoliubov coefficients.

Similar to \eqref{LF:mesonic:operators:commutator} in the LF case, here we again postulate that the mesonic
annihilation and creation operators obey
the canonical commutation relations:
\beq
\left[m_n(P),m^\dagger_m(P')\right]=2\pi\,\delta_{n m}\,\delta(P-P').
\label{Equal:time:m:mdagger:commutation}
\eeq

To satisfy these commutation relations, the Bogoliubov functions $\varphi^n_\pm$ must obey the following
orthogonality and completeness conditions\footnote{We emphasize that the BGWFs are normalized differently from our preceding work~\cite{Jia:2018qee}.}:
\bseq
\begin{align}
&\int_{-\infty}^{+\infty}\frac{dp}{2\pi}\left[\varphi_+^n(p,P)\,\varphi_+^m(p,P)-\varphi_-^n(p,P)\,\varphi_-^m(p,P)\right]=\delta^{nm},
\\
&\int_{-\infty}^{+\infty}{dp}\left[\varphi_+^n(p,P)\,\varphi_-^m(p-P,-P)-\varphi_-^n(p,P)\,\varphi_+^m(p-P,-P)\right]=0,
\\
&\sum_{n=0}^\infty\left[\varphi_+^n(p,P)\,\varphi_+^n(q,P)-\varphi_-^n(p-P,-P)\,\varphi_-^n(q-P,-P)\right]={2\pi}{\delta(p-q)},
\\
&\sum_{n=0}^\infty\left[\varphi_+^n(p,P)\,\varphi_-^n(q,P)-\varphi_-^n(p-P,-P)\,\varphi_+^n(q-P,-P)\right]=0.
\end{align}
\eseq
Note the relative minus sign is reminiscent of the
characteristic of the Bogoliubov transformation~\cite{Kalashnikova:2001df}.

Substituting \eqref{M:expressed:in:mn} into \eqref{Hamiltonian:re:split:three:pieces}, we attempt to
put the $\mathbb{H}^{(0)}$ into the diagonalized form, which describes
an infinite towers of noninteracting mesons:
\beq
\mathbb{H}^{(0)}= \sum_n \int_{-\infty}^\infty\frac{dP}{2\pi}P_n^0 m_n^\dagger(P)m_n(P),
\label{H0:ET:diagonal:form}
\eeq
where $P_n^0 =\sqrt{\mu_n^2+P^2}$.

We define the mesonic vacuum state $\vert \Omega \rangle$ by the condition
$m_n(P)\vert \Omega \rangle = 0$,
for all $n$ and $P$. Consequently,
a single $n$th excited mesonic state can be constructed via
\beq
\vert P \rangle = \sqrt{2P_n^0}m^\dagger_n(P) | \Omega \rangle.
\eeq
Note the mesonic vacuum state $\vert \Omega \rangle$ differs from the quark vacuum $|0\rangle$ in the equal-time quantization,
and is highly nontrivial.

To achieve the intended diagonal form \eqref{H0:ET:diagonal:form}, one must enforce that all the off-diagonal terms in $\mathbb{H}^{(0)}$ cancel,
which in turn imposes the following constraints on the Bogliubov coefficient functions $\varphi_{n}^{\pm}(p, P)$:
\begin{align}
 \left[\mathcal{E}(p)+\mathcal{E}(P\!-\!p) \mp P_{n}^{0}\right] \varphi_{n}^{\pm}(p, P)=\lambda \dashint_{-\infty}^{+\infty} \frac{d k}{(p\!-\!k)^{2}} \left[C(p, k, P) \varphi_{n}^{\pm}(k, P)\!-\!S(p, k, P) \varphi_{n}^{\mp}(k, P)\right],
 \label{Final:BG:bound-state:eqs}
 \end{align}
with
\bsq
\begin{align}
 &C(p, k, P)=\cos \frac{\theta(p)-\theta(k)}{2} \cos \frac{\theta(P-p)-\theta(P-k)}{2},
 \\
 &S(p, k, P)=\sin \frac{\theta(p)-\theta(k)}{2} \sin \frac{\theta(P-p)-\theta(P-k)}{2}.
\end{align}
 \esq

Equation~(\ref{Final:BG:bound-state:eqs}) is the coupled bound-state equations in the 't Hooft model in equal-time quantization,
first derived by Bars and Green back in 1978~\cite{Bars:1977ud}.
For this reason, this equation will be referred to as the Bars-Green equation.
Consequently, the Bogoliubov coefficient functions $\varphi^n_\pm$ are interpreted as
the forward/backward-moving bound-state wave functions, or simply called Bars-Green wave functions.
The BGWFs of a flavor-neutral meson are subject to the following constraints from discrete symmetries, such
as the parity and charge-conjugation symmetries:
\bseq
 \begin{eqnarray}
\varphi^{\pm}_n(-p,P)&=(-1)^n\varphi^{\pm}_n(p,P)\quad\quad &\mathrm{P},
\\
\varphi^{\pm}_n(P-p,P)&=(-1)^n\varphi^{\pm}_n(p,P)\quad\quad &\mathrm{C},
\\
\varphi^{\pm}_n(p-P,-P)&=\varphi^{\pm}_n(p,P)\phantom{(-1)^n}\quad\quad &\mathrm{CP}.
 \end{eqnarray}
 \label{eq:BGWF_cpt}
 \eseq

Apparently, the Bars-Green equations in (\ref{Final:BG:bound-state:eqs}) are much more complicated than it's counterpart in IMF, the 't Hooft equation \eqref{eq:tHooft_eq}.
This complication can be largely attributed to the nontrivial vacuum structures in equal-time quantization.
As indicated in \eqref{M:expressed:in:mn}, a meson may be created out of the vacuum by annihilating a pair of quark and antiquark, due to the emergence of the backward-moving
component of the BGWF $\varphi^-$.
It is worth emphasizing that, a crucial virtue of Bars-Green equations is that it preserves the Poincar\'{e} invariance
in physical sector in a highly nontrivial way.
A specific consequence of Poincar\'{e} invariance is that,
when the meson is viewed in the IMF, that is, in the $P\to \infty$ limit,
one would still obtain the identical mesonic mass spectra.
In the $P\to \infty$  limit, by relabeling the quark momentum by $p=x P$,  one readily shows that that
 \begin{align}
   \theta(xP)\approx \epsilon(x)\frac{\pi}{2},\quad \tan \theta(xP) \approx \frac{xP}{m},\quad
   \mathcal{E}(xP) = |x|P+\frac{m^2-2\lambda}{2|x|P}+\mathcal{O}\left(\frac{1}{P^2}\right),
 \label{eq:BGasy}
 \end{align}
 where $\epsilon(x)$ denotes the sign function.  Substituting the above asymptotic behavior \eqref{eq:BGasy} into the Bars-Green equation
 \eqref{Final:BG:bound-state:eqs}, and only retaining the $\mathcal{O}(1/P)$ pieces,  one finds that
 \begin{align}
 \lim_{P\to\infty}\sqrt{\frac{P}{2\pi}}\varphi_n^+(xP,P) = \Theta(x)\Theta(1-x)\varphi_n(x),\qquad \lim_{P\to\infty}\varphi_n^-(xP,P) = 0.
 \label{eq:BGasy2}
 \end{align}
Therefore, in the infinite momentum limit, the forward-moving BGWF $\varphi_n^+$ is approaching the LCWF, while the backward-moving BGWF $\varphi_n^-$ fades away.
 Consequently the Bars-Green equation reduces to the 't Hooft equation \eqref{eq:tHooft_eq}.

The last piece $\mathbb{V}$ in the Hamiltonian \eqref{Hamiltonian:re:split:three:pieces} scales as ${\cal O}(1/\sqrt{N_c})$,  which entails
the integration of the triple product of $m$ and $m^\dagger$ operators.
Its full expressions are quite lengthy, and the part that is responsible for the quark quasi-GPD is collected in
Appendix~\ref{appendix:B}.
Here we just exhibit one typical term in $\V$:
\begin{align}
  \notag \V=&\frac{\lambda}{\sqrt{N_c}}\sum_{n_1,n_2,n_3}\int \frac{dk_1dk_2dk_3dk_4dk_5}{(2\pi)^3}\Bigg[-\cos \frac{\theta \left(k_1\right)-\theta \left(k_4\right)}{2}\sin \frac{\theta \left(k_2\right)-\theta \left(k_3\right)}{2} \frac{\delta \left(-k_1+k_2-k_3+k_4\right)}{\left(k_2-k_3\right)^2} \\
     &\times m_{n_1}^{\dagger }\left(k_1-k_2\right)m_{n_2}^{\dagger }\left(k_3-k_5\right)m_{n_3}\left(k_4-k_5\right)\varphi_{n_1}^-\left(k_1,k_1-k_2\right)\varphi_{n_2}^-
     \left(k_3,k_3-k_5\right)\varphi_{n_3}^-\left(k_4,k_4-k_5\right)+\cdots\Bigg].
\label{eq:V_incomplete}
\end{align}
{These operators induce the three-meson interaction vertex, which represents an
${\cal O}(1/\sqrt{N_c})$ effect}.

\section{Light-Cone GPD in 't Hooft model}
\label{Sec:LCGPD:derivation}

Employing diagrammatic approach, the light-cone GPD of a charged meson in the 't Hooft model was considered by Burkadrt in 2000~\cite{Burkardt:2000uu}.
In the following, we will utilize the Hamiltonian approach to derive the functional form of the light-cone GPD
of a flavor-neutral meson in the 't Hooft model.

 \subsection{$1/N_c$ expansion of the quark bilinear operator and meson states in light-front quantization}

 We first consider the nonlocal quark bilinear operator in the light-cone GPD  defined in \eqref{eq:lcgpd_s0}.
 Working with the light-cone gauge $A^{+,a}=0$, the gauge link in \eqref{eq:lcgpd_s0} shrinks to the unit operator.
 Following the bosonization procedure outlined in Sec.~\ref{H:approach:LF:quantization},
 we can recast the quark bilinear with lightlike separation in \eqref{eq:lcgpd_s0} in terms of the mesonic annihilation and creation operators.
 We split the bosonzied quark bilinear operator into three pieces:
  \beq
 \bar{\psi}\left(-\frac{\eta^-}{2}\right)\gamma^{+}\psi\left(\frac{\eta^-}{2}\right) =
 \psi^\dagger_{R}\left(-\frac{\eta^-}{2}\right)\psi_{R}\left(\frac{\eta^-}{2}\right) = {\OO}_1(\eta^-)  +\OO_{1/2}(\eta^-) + {\OO}_0(\eta^-) + \cdots,
\eeq
where
\bseq
\beq
\OO_1(\eta^-) =  N_c \int {dk^+\over 2\pi}\; e^{ik^+\eta^-},\qqquad\qqquad\qqquad\qqquad\qqquad\qqquad\qqquad\quad
\eeq
\begin{align}
 \notag & \OO_{1/2}(\eta^-) =\sqrt{\frac{N_c}{2\pi}}\sum_{r}\int \frac{dk_1^+dk_2^+}{2\pi\sqrt{k_1^++k_2^+}}\Bigg[e^{i \eta^- \left(k_2^+\!-\!k_1^+\right)/2} m_{r}^{\dagger }\!\left(k_1^+\!+\!k_2^+\right) \varphi_{r}\! \left(\frac{k_1^+}{k_1^+\!+\!k_2^+}\right) +(k_1^+ \leftrightarrow k_2^+) \Bigg],
\end{align}
\begin{align}
  \notag &\OO_0(\eta^-)=\int\frac{dk_1^+ dk_2^+ dk_3^+}{4\pi^2}\Bigg[
  e^{-i \eta^-\left(k_1^+\!+\!k_2^+\right)/2} \frac{ m_{n_1}^{\dagger }\left(k_1^+\!+\!k_3^+\right)m_{n_2}\!\left(k_2^+\!+\!k_3^+\right) }{\sqrt{k_1^+\!+\!k_3^+}\sqrt{k_2^+\!+\!k_3^+}}\varphi_{n_1}\! \left(\frac{k_1^+}{k_1^+\!+\!k_3^+}\right) \varphi_{n_2}\! \left(\frac{k_2^+}{k_2^+\!+\!k_3^+}\right)
  \\
  &\qqquad\qquad -e^{i\eta^-\left(k_1^+\!+\!k_2^+\right)/2} \frac{ m_{n_1}^{\dagger }\!\left(k_2^+\!+\!k_3^+\right)m_{n_2}\!\left(k_1^+\!+\!k_3^+\right)}{\sqrt{k_1^+\!+\!k_3^+}\sqrt{k_2^+\!+\!k_3^+}}  \varphi_{n_1}\! \left(\frac{k_3^+}{k_2^+\!+\!k_3^+}\right) \varphi_{n_2}\! \left(\frac{k_3^+}{k_1^+\!+\!k_3^+}\right)
  \Bigg],
  \end{align}
\label{eq:bilinear_IMF}
\eseq
where the subscripts of $\OO$ indicate the power of $N_c$ affiliated with the respective mesonic operators.

The leading-color operator $\OO_1(\eta^-)$ becomes a unit operator. When sandwiched between the antinial and final mesonic states carrying different momenta, the corresponding
matrix element vanishes. In the forward limit, the matrix element involving $\OO_1$ is affiliated with the disconnected part, and can be subtracted according to the definition of light-cone PDF. Thus, we do not need consider the contribution of  $\OO_1$.

 $\OO_{1/2}(\eta^-)$ involves integrations of a single $m_n$ ($m_n^\dagger$) operator,  while $\OO_0(\eta^-)$
 involves integrations of $m_{n_1}^\dagger m_{n_2}$. Recalling the definition of the single meson state in the light-front quantization in \eqref{single:meson:LF:quantization},
 one immediately sees that the matrix element involving $\OO_0(\eta^-)$ yields a nonvanishing ${\cal O}(N_c^0)$ contribution.

 At first sight, the matrix element involving $\OO_{1/2}(\eta^-)$ is expected to vanish, which originates from the vacuum matrix element of the product of an odd number of mesonic
 annihilation and creation operators. Nevertheless, one has to caution that the meson states defined in \eqref{eq:lcgpd_s0} are the eigenstates of full light-front Hamiltonian
$\mathbb{H}_{\mathrm{LF}}$ in \eqref{eq:HLF_full}, rather than the eigenstates of the free mesonic Hamiltonian $\mathbb{H}_{\mathrm{LF}}^{(0)}$,
exemplified by the single meson state in \eqref{single:meson:LF:quantization}.
Including the first-order quantum mechanical perturbation, the physical meson state can be expressed as
\begin{align}
 \vert P\rangle' &= \vert P\rangle + \frac{1}{P^--\Hlfz + i\epsilon}\Vlf\vert P \rangle + \cdots.
\label{LF:Lippmann-Schwinger:meson:state:expansion}
 \end{align}
 $\Vlf$ represents the $\mathcal{O}(N_c^{-1/2})$ interacting Hamiltonian that induces a meson to transition into two mesons, as exemplified by \eqref{eq:V_incomplete}.
 Therefore, the second term in the right-hand side of \eqref{LF:Lippmann-Schwinger:meson:state:expansion} represents a
 two-meson higher Fock state, which is suppressed by
 ${\mathcal O}(1/\sqrt{N_c})$  with respect to the leading Fock component.

In fact, the matrix element of $\OO_{1/2}(\eta^-)$ yields a net ${\mathcal O}(N_c^0)$ contribution from
the higher Fock components of the initial- or final-state mesons.
This piece of contribution to the light-cone GPD should be supplemented to the matrix element of $\OO_{0}(\eta^-)$, both of which scale as ${\mathcal O}(N_c^0)$.

\subsection{Deriving the functional form of light-cone GPD}
\label{sec:der_lc_gpd}

Substituting the bosonized quark bilinear \eqref{eq:bilinear_IMF} and  \eqref{LF:Lippmann-Schwinger:meson:state:expansion} into \eqref{eq:lcgpd_s0},
we find that the leading $\mathcal{O}(N_c^0)$ contribution to the light-cone GPD/quasi-GPD consists of two parts
\beq
H(x,\xi)= H^{(0)}(x,\xi)+H^{(1)}(x,\xi),
\eeq
where
\bseq
\begin{align}
&H^{(0)}(x,\xi) =\int\frac{d\eta^-}{2\pi}e^{ixP^+\eta^-} \left\langle P+\frac{\varDelta}{2} \right|\OO_0(\eta^-) \left| P-\frac{\varDelta}{2} \right \rangle,
\label{eq:GPD_IMF_expnd_V0}
\\
\notag &H^{(1)}(x,\xi) = \int\frac{d\eta^-}{4\pi}e^{ixP^+\eta^-} \left\{ \left\langle P+\frac{\varDelta}{2} \left| \Vlf\,\frac{1}{P^-+\frac{\varDelta^-}{2}-\Hlfz-i\epsilon} \OO_{1/2}(\eta^-)\right| P-\frac{\varDelta}{2} \right\rangle \right.
\\
&\qqquad\qqquad\; \left. + \left\langle P+\frac{\varDelta}{2} \left|\OO_{1/2}(\eta^-)\frac{1}{P^- - \frac{\varDelta^-}{2}-\Hlfz+i\epsilon}\Vlf \right| P-\frac{\varDelta}{2}
\right\rangle  \right\}.
\label{eq:GPD_IMF_expnd_V1}
\end{align}
\eseq
The superscript in $H(x,\xi)$ indicates the order of the Fock component of the mesonic state which contributes to the light-cone GPD.

Computation of \eqref{eq:GPD_IMF_expnd_V0} is straightforward. For the $n$th excited mesonic state, we have
\begin{align}
 \notag H^{(0)}_n(x,\xi)=&\;\Theta(x-\xi)\Theta(1-x)\Theta(\xi+x)\varphi_n\left(\frac{x-\xi}{1-\xi}\right)\varphi_n\left(\frac{x+\xi}{1+\xi}\right),
 \\
 &-\!\Theta(1+x)\Theta(-x-\xi)\Theta(\xi-x)\varphi_n\left(\frac{-x-\xi}{1-\xi}\right)\varphi_n\left(\frac{-x+\xi}{1+\xi}\right).
\label{eq:LCGPD_exp_cmpct_dglap}
\end{align}

Note that $H^{(0)}$ is only nonvanishing in the so-called DGLAP region $\xi<|x|<1$, where the quark light-cone GPD is interpreted as {the amplitude of emission an (anti)quark from the meson then being absorbed again by the meson~\cite{Diehl:2003ny}.}

As pointed out by Burkadrt~\cite{Burkardt:2000uu}, the $H^{(0)}(x,\xi)$ in the DGLAP region
exactly saturates the positivity bound  \eqref{eq:positivity}. This may be attributed to a peculiarity of {\QCDtw} that there is no transverse degrees of freedom.

In contrast to \eqref{eq:GPD_IMF_expnd_V0}, computation of $H^{(1)}_n(x,\xi)$ is more involved and needs some explanations.
We take the first term in \eqref{eq:GPD_IMF_expnd_V1} as a concrete example to demonstrate the derivation.
Inserting a unit operator between $\Vlf$ and the energy denominator, we obtain
\begin{align}
&\left\langle P+\frac{\varDelta}{2} \left| \Vlf\,\frac{1}{P^-+\frac{\varDelta^-}{2}-\Hlfz-i\epsilon} \OO_{1/2}(\eta^-)\right| P-\frac{\varDelta}{2} \right \rangle
\nn\\
=& {1\over 2!}\sum_{n_1,n_2}\int\frac{dq_1^+dq_2^{+}}{\left(2\pi\right)^22q_1^+2q_2^{+}}
\left\langle P+\frac{\varDelta}{2},n \left| \Vlf \bigg\vert q_1,n_1;q_2,n_2\bigg\rangle\bigg\langle q_1,n_1;q_2,n_2\bigg\vert \OO_{1/2}(\eta^-)\right| P-\frac{\varDelta}{2},n \right \rangle \,\frac{1}{P^-+\frac{\varDelta^-}{2}-q_1^- - q_2^{-}},
\label{eq:I_insert}
\end{align}
where $q_{1,2}^- = \mu_{n_{1,2}}^2/(2q_{1,2}^+)$ with $\mu_{n_{1,2}}$ denoting the masses of the $n_{1,2}$th mesonic state.
We have utilized the fact that when the unit operator is expanded in the basis of all possible light-front energy eigenstates, only those spanned by all two-meson intermediate states can render a nonvanishing contribution:
\beq
 \mathbb{I} = \frac{1}{2!}\sum_{n_1,n_2}\int\frac{d q_1^+ dq_2^+} {\left(2\pi\right)^2 2q_1^+2 q_2^{+}} \vert q_1,n_1;q_2,n_2\rangle\langle q_1,n_1;q_2,n_2\vert + \cdots.
\eeq

The first matrix element in the integrand in \eqref{eq:I_insert} can be expressed as
\beq
\langle P+{\varDelta\over 2},n \vert \Vlf \vert  q_1,n_1; q_2,n_2\rangle \equiv {2\pi}\delta\left(P^+ +  {\varDelta^+ \over 2 } - q_1^+ - q_2^{+} \right)
 \varGamma_{n,n_1,n_2} \left(\frac{q_1^+}{P^+ + \varDelta^+/2},\frac{q_2^+}{P^+ + \varDelta^+/2}\right).
\label{meson:transition:to:two:meson:matrix:element}
\eeq
The three-meson vertex function $\varGamma$ in light-front quantization was first introduced by
Callan {\it et al.}~\cite{Callan:1975ps}, which can be expressed as the convolution of three meson's LCWFs.
Its explicit expression is given in Appendix~\ref{appendix:A}.

For the second matrix element in the integrand in \eqref{eq:I_insert}, it is the $m^\dagger$ component of the
operator $\OO_{1/2}(\eta^-)$ that yields a nonvanishing contribution. Using Bose symmetry and the commutation relation
\eqref{LF:mesonic:operators:commutator},
one can decompose this matrix element into the product of two clusters:
\begin{align}
  \notag &\bigg\langle q,n_1;q_2,n_2\bigg\vert\mathbb{O}_{1/2}(\eta^-)\bigg\vert P-\frac{\varDelta}{2},n\bigg\rangle
  \\
=& \bigg\langle q_1, n_1\bigg\vert P-\frac{\varDelta}{2},n\bigg\rangle
 \bigg\langle q_2,n_2\bigg\vert\mathbb{O}_{1/2}(\eta^-)\bigg\vert 0 \bigg\rangle + (q_1\leftrightarrow q_2,\; n_1\leftrightarrow n_2).
\label{Reduction:of:O1/2:matrix:element}
\end{align}
Apparently these two terms yield identical contributions to \eqref{eq:I_insert} upon integrating over $q_1$ and $q_2$ and summing over $n_1$ and $n_2$.

Equation~\eqref{Reduction:of:O1/2:matrix:element} involves the inner product between two single meson states,
which is subject to the orthogonality condition:
\begin{align}
  \bigg\langle q_1,n_1\bigg\vert P-\frac{\varDelta}{2},n\bigg\rangle =  (2\pi) (2q_1^+) \delta_{n, n_1} \delta\left(P^+-\frac{\varDelta^+}{2}-q_1^+\right).
\label{eq:LF:state:orth}
\end{align}
In addition, the momentum conservation in \eqref{meson:transition:to:two:meson:matrix:element} enforces that $q_1^+ + q_2^{+}= P^+ +\varDelta^+/2$.
Thus, combining these two $\delta$ functions  can uniquely determine $q_1^+=P^+-\varDelta^+/2=(1-\xi)P^+$,
$q_2^+=\varDelta^+=2\xi P^+$, as well as $n_1=n$.
Consequently, one can make the following substitution for energy denominator in \eqref{eq:I_insert:equal:time:quantization}:
\begin{align}
\frac{1}{P^-+\frac{\varDelta^-}{2} - q_1^- - q_2^-} \to  \frac{4\xi P^+}{t - \mu_{n_2}^2}.
\end{align}
where $t \equiv \varDelta^2=\frac{4\mu_n^2\xi^2}{\xi^2-1}$, as given in \eqref{eq:t_xi_lc}.

It is convenient to introduce an auxiliary function $\varPsi$:
 \begin{align}
 & \varPsi_{s}\left(x,\frac{q^+}{P^+}\right) \equiv\int\frac{d\eta^-}{4\pi}\,e^{ix P^+\eta^-}{{\bra{q,s}\OO_{1/2}(\eta^-)\ket{0}}},
\label{Def:Psi_function:LF:quantization}
\end{align}
with the single mesonic state labeled by the principle quantum number $s$ and the light-cone momentum $q^+$.

After some straightforward manipulation, we are able to express $H^{(1)}_n(x,\xi)$
in \eqref{eq:GPD_IMF_expnd_V1} as
\beq
H^{(1)}_n(x,\xi)= \sum_{r}{\frac{1}{t-\mu_{r}^2}}\left[\Theta(\xi)
 \varGamma_{n,n,r}\left(\frac{1\!-\!\xi}{1\!+\!\xi},\frac{2\xi}{1\!+\!\xi}\right)
 \varPsi_{r}\left(x, 2\xi\right)+\Theta(-\xi)
 \varGamma_{n,n,r}\left(\frac{1\!+\!\xi}{1\!-\!\xi},\frac{2\xi}{\xi\!-\!1}\right)
 \varPsi_{r}\left(x, -2\xi\right)\right],
\label{eq:H1_IMF_exp}
\eeq
where $n_2$ has been relabeled by $r$.

One can readily work out the closed form for the auxiliary function:
\begin{align}
  \varPsi_{r}(x,2\xi) =\sqrt{\frac{N_c}{4\pi}}\Theta(\xi+x)\Theta(\xi-x)\varphi_{r}\left(\frac{x+\xi}{2\xi}\right).
\label{Psi:LC:analytical:expression}
\end{align}

Substituting \eqref{Psi:LC:analytical:expression} and $\varGamma$'s explicit form \eqref{eq:3msnvrtx_LCWF} into \eqref{eq:H1_IMF_exp},
we finally derive an analytic expression for $H_n^{(1)}$:
\begin{align}
H^{(1)}_n(x,\xi)=&\;\Theta(\xi)\Theta(\xi+x)\Theta(\xi-x)2\lambda\int_{0}^{2\xi}dw\int_{0}^{1-\xi}dy\frac{1}{(y+w)^{2}}\sum_{r}\frac{1}
{t-\mu_{r}^{2}}\varphi_{r}\left(\frac{w}{2\xi}\right)\left[\varphi_{r}\left(\frac{\xi-x}{2\xi}\right)-\varphi_{r}\left(\frac{\xi+x}{2\xi}\right)\right]
\nn\\
&\quad \times\left[\varphi_n\left(\frac{2\xi-w}{1+\xi}\right)-\varphi_n\left(\frac{2\xi+y}{1+\xi}\right)\right]\varphi_n\left(\frac{y}{1-\xi}\right) + (\xi \rightarrow -\xi).
\label{eq:LCGPD_exp_cmpct_erbl}
 \end{align}
Note that $H_n^{(1)}$ is only nonvanishing in the so-called ERBL region ($|x|<\xi$),
which can be interpreted as the amplitude of extracting a quark-antiquark pair from the hadron~\cite{Diehl:2003ny}.

It is straightforward to check that the light-cone GPD is an even functions of $\xi$. One can also verify that the light-cone GPD is an odd function of $x$,
reflecting the charge-conjugation symmetry of a flavor-neutral meson~\cite{Belitsky:2005qn}.

Equations~\eqref{eq:LCGPD_exp_cmpct_dglap} and \eqref{eq:LCGPD_exp_cmpct_erbl} constitute two of the most important equations in this work, which
express the light-cone GPD of a flavor-neutral meson in the 't Hooft model in a closed form.
Our result is supplementary to the expression of the light-cone GPD for a flavored meson in the 't Hooft model~\cite{Burkardt:2000uu}.
It may be worth emphasizing that, the light-cone PDF in the 't Hooft model is exceedingly simple, which is expressed in term of a single meson's LCWF.
In contrast, the light-cone GPD $H_n^{(1)}$ in this model already contains very rich information,  which receives contribution from an infinite tower of excited
mesonic states. In the large-$N_c$ limit of ${\rm QCD}_4$, the light-cone GPD of a meson is expected to also involve the sum of infinite towers of excited mesons.
It is hard to imagine how complicated the light-cone GPD in realistic QCD might look.

\section{Quasi-GPD in 't Hooft model}
\label{Sec:Quasi:GPD:derivation}

In this section, we will utilize the Hamiltonian approach to derive the functional form of the quasi-GPD of a flavor-neutral meson in the 't Hooft model.
A significant difference is that we turn to the equal-time quantization. Otherwise
the derivation is parallel to the light-cone GPD case as outlined in Sec.~\ref{Sec:LCGPD:derivation}.

\subsection{$1/N_c$ expansion of the quark bilinear operator and meson states in equal-time quantization}

 We begin with the nonlocal quark bilinear operator in the quasi-GPD  defined in \eqref{eq:quasigpd_s0}.
 Imposing the axial gauge $A^{z,a}=0$, the gauge link disappears in \eqref{eq:quasigpd_s0}.
 Following the bosonization procedure outlined in Sec.~\ref{H:approach:ET:quantization},
 we can recast the quark bilinear with purely spacelike separation in \eqref{eq:quasigpd_s0} in terms of the mesonic annihilation and creation operators.
 We break the bosonzied quark bilinear operator into three pieces:
\beq
\overline{\psi}\left(-\frac{z}{2}\right)\gamma^z\psi\left(\frac{z}{2}\right)=\widetilde{\OO}_1(z)  + \widetilde{\OO}_{1/2}(z) + \widetilde{\OO}_{0}(z) + \cdots,
\label{nonlocal:bilinear:equal:time:split}
\eeq
where
\bsq
\beq
\widetilde{\OO}_1(z) =  \frac{N_c}{2\pi} \int dk\, \sin \theta(k)\,e^{-ikz},\qqquad\qqquad\qqquad\qqquad\qqquad\qqquad\qqquad\qqquad\qqquad\quad
\eeq
\begin{align}
&\widetilde{\OO}_{1/2}(z)= -\frac{\sqrt{N_c}}{4\pi^2}\int\! dk_1dk_2e^{iz({k_1\!+\!k_2})/2}\Bigg\{\sum_{n_1} \Big[m_{n_1}^{\dagger }\left(k_1\!-\!k_2\right) \varphi_{n_1}^-\left(k_1,k_1\!-\!k_2\right)\!+\!m_{n_1}\left(k_2\!-\!k_1\right) \varphi_{n_1}^-\left(k_2,k_2\!-\!k_1\right)
\nn \\
&\qqquad-m_{n_1}^{\dagger }\left(k_1\!-\!k_2\right) \varphi_{n_1}^+\left(k_1,k_1\!-\!k_2\right)\!-\!m_{n_1}\left(k_2\!-\!k_1\right) \varphi_{n_1}^+\left(k_2,k_2\!-\!k_1\right)\Big]\cos \frac{\theta \left(k_1\right)\!+\!\theta \left(k_2\right)}{2},
\end{align}
\begin{align}
  \widetilde{\OO}_0(z)=\frac{1}{8\pi^3}&\int dk_1 dk_3 dk_3e^{iz(k_1+k_2)/2}\sum_{n_1,n_2}\Big[
      \varphi_{n_1}^-\left(k_2,k_2-k_3\right) \varphi_{n_2}^-\left(k_1,k_1-k_3\right) m_{n_1}\left(k_2-k_3\right)m_{n_2}^{\dagger }\left(k_1-k_3\right) \notag \\
    &+\varphi_{n_1}^-\left(k_3,k_3-k_1\right) \varphi_{n_2}^-\left(k_3,k_3-k_2\right) m_{n_1}\left(k_3-k_1\right)m_{n_2}^{\dagger }\left(k_3-k_2\right) \notag \\
    &-\varphi_{n_2}^-\left(k_3,k_3-k_2\right) \varphi_{n_1}^+\left(k_1,k_1-k_3\right) m_{n_1}^{\dagger }\left(k_1-k_3\right)m_{n_2}^{\dagger }\left(k_3-k_2\right) \notag \\
    &-\varphi_{n_2}^-\left(k_1,k_1-k_3\right) \varphi_{n_1}^+\left(k_3,k_3-k_2\right) m_{n_1}^{\dagger }\left(k_3-k_2\right)m_{n_2}^{\dagger }\left(k_1-k_3\right) \notag \\
    &-\varphi_{n_1}^-\left(k_3,k_3-k_1\right) \varphi_{n_2}^+\left(k_2,k_2-k_3\right)m_{n_1}\left(k_3-k_1\right)m_{n_2}\left(k_2-k_3\right)  \notag \\
    &-\varphi_{n_1}^-\left(k_2,k_2-k_3\right) \varphi_{n_2}^+\left(k_3,k_3-k_1\right)m_{n_1}\left(k_2-k_3\right)m_{n_2}\left(k_3-k_1\right)  \notag \\
    &+\varphi_{n_1}^+\left(k_1,k_1-k_3\right) \varphi_{n_2}^+\left(k_2,k_2-k_3\right) m_{n_1}^{\dagger }\left(k_1-k_3\right)m_{n_2}\left(k_2-k_3\right) \notag \\
    &+\varphi_{n_1}^+\left(k_3,k_3-k_2\right) \varphi_{n_2}^+\left(k_3,k_3-k_1\right) m_{n_1}^{\dagger }\left(k_3-k_2\right)m_{n_2}\left(k_3-k_1\right)\Big]\sin\frac{\theta(k_1)+\theta(k_2)}{2},
  \end{align}
\label{eq:bilinear:equal:time}
\esq
where the subscript in $\widetilde{\OO}(z)$ indicates the power of $N_c$ affiliated with the respective mesonic operators.

The leading-color operator $\widetilde{\OO}_1(z)$ is proportional to a unit operator and does not contribute to the quasi-GPD; therefore we will simply discard it.
$\widetilde{\OO}_{1/2}(z)$ involves integrations of a single $m_n$ ($m_n^\dagger$) operator,  while $\widetilde{\OO}_{0}(z)$
involves integrations of $m_{n_1}^\dagger m_{n_2}$.

It is apparent that the matrix element involving $\widetilde{\OO}_{0}(z)$ yields a nonvanishing ${\cal O}(N_c^0)$  contribution.
It is worth emphasizing that, similar to the light-cone GPD case,  the $\widetilde{\OO}_{1/2}(z)$ operator also makes a nonvanishing
${\cal O}(N_c^0)$  contribution to the matrix element,
provided that the next-to-leading-order Fock component of a physical mesonic state is considered.
Incorporating the first-order quantum mechanical perturbation, the physical meson state can be expressed as
\begin{align}
 \vert P\rangle'& = \vert P \rangle + \frac{1}{P^0-\Hz + i\epsilon}\V\vert P \rangle + \cdots,
\label{Lippmann:Schwinger:eq:equal:time}
 \end{align}
 where $\V$ represents the $\mathcal{O}(N_c^{-1/2})$ interacting Hamiltonian that induces a meson to
fluctuating into two mesons, which is introduced in \eqref{Hamiltonian:re:split:three:pieces}.
The second term in the right-hand side of \eqref{Lippmann:Schwinger:eq:equal:time} thus represents
a two-meson Fock component, which is suppressed by ${\mathcal O}(1/\sqrt{N_c})$  with respect to the leading Fock component.
Consequently,
the matrix element of $\widetilde{\OO}_{1/2}(z)$ yields an ${\mathcal O}(N_c^0)$ contribution from
the higher Fock component of the initial meson state or final meson state.
Therefore, both the $\widetilde{\OO}_{0}(z)$ and $\widetilde{\OO}_{1/2}(z)$ operators result in
an ${\mathcal O}(N_c^0)$ contribution to the quasi-GPD of a meson.

\subsection{Deriving the functional form of quasi-GPD}

Plugging the bosonized quark bilinear \eqref{eq:bilinear:equal:time} and \eqref{Lippmann:Schwinger:eq:equal:time} into \eqref{eq:quasigpd_s0},
we find that the leading $\mathcal{O}(N_c^0)$ contribution to the quasi-GPD contains two parts:
\beq
\widetilde{H}(x,\xi,P^z)= \widetilde{H}^{(0)}(x,\xi,P^z)+\widetilde{H}^{(1)}(x,\xi,P^z),
\label{breaking:quasi:GPD:two:pieces}
\eeq
where
\bsq
\begin{align}
&\widetilde{H}^{(0)}(x,\xi,P^z) =\int\frac{dz}{2\pi}e^{-ixP  z}{\left\langle P+\frac{\varDelta}{2} \right| \widetilde{\OO}_0(z) \left| P-\frac{\varDelta}{2} \right \rangle},
\label{eq:GPD_FMF_expnd_V0}
\\
\notag  &\widetilde{H}^{(1)}(x,\xi,P^z) = \int\frac{dz}{4\pi}e^{-ix P z}
\left\langle P+\frac{\varDelta}{2} \left| \V\,\frac{1}{P^0+\frac{\varDelta^0}{2}-\Hz-i\epsilon}   \widetilde{\OO}_{1/2}(z)\right| P-\frac{\varDelta}{2} \right \rangle
\\
&\qqquad\qqquad\quad +{\left\langle P+\frac{\varDelta}{2} \left| \widetilde{\OO}_{1/2}(z) \frac{1}{P^0-
\frac{\varDelta^0}{2}-\Hz+i\epsilon}\V \right| P-\frac{\varDelta}{2} \right \rangle}.
\label{eq:GPD_FMF_expnd_V1}
\end{align}
\label{quasi:GPD:two:parts:def}
\esq
The superscript in $\widetilde{H}$ indicates the order of the respective Fock component of the mesonic state that contributes to the quasi-GPD.

Computation of \eqref{eq:GPD_FMF_expnd_V0} is straightforward. Repeatedly using the commutation relation in \eqref{Equal:time:m:mdagger:commutation},
we find that for the $n$th excited mesonic state,
\begin{align}
 \widetilde{H}_n^{(0)}(x,\xi,P^z) =&\frac{\sqrt{E_1 E_2}}{2 \pi } \sin \frac{\theta \left(k_1\right)+\theta \left(k_2\right)}{2} \left[
\varphi_n^+\left(k_1,P_1\right) \varphi_n^+\left(k_2,P_2\right)+\varphi_n^+\left(-k_2,P_1\right) \varphi_n^+\left(-k_1,P_2\right)
\right.
\nn \\
& \left.  + \varphi_n^-\left(k_1,P_1\right) \varphi_n^-\left(k_2,P_2\right)+\varphi_n^-\left(-k_2,P_1\right) \varphi_n^-\left(-k_1,P_2\right)
\right],
\label{quasi:GPD:tildeH0:expression}
\end{align}
where
\beq
P_{1}=(1-\xi)P^z,\quad P_{2}=(1+\xi)P^z, \quad k_{1}=(x-\xi)P^z,\quad \quad k_{2}=(x+\xi)P^z, \quad {E}_{1,2} =\sqrt{\mu_n^2+P_{1,2}^2}.
\eeq

The computation of $\widetilde{H}^{(1)}$ is more involved, but the strategy is quite parallel to
the computation of $H^{(1)}$ as expounded in Sec.~\ref{sec:der_lc_gpd}.
We demonstrate the derivation by taking the first term in \eqref{eq:GPD_FMF_expnd_V1} as an concrete example.
Inserting a unit operator between $\V$ and the energy denominator, we find
\begin{align}
& \left\langle P+\frac{\varDelta}{2}, n \left| \V\,\frac{1}{P^0+\frac{\varDelta^0}{2}-\Hz-i\epsilon}
\widetilde{\OO}_{1/2}(z)\right| P-\frac{\varDelta}{2}, n \right \rangle
\nn \\
=&   \frac{1}{2!}\sum_{n_1,n_2}\int\frac{dq_1 dq_2}{\left(2\pi\right)^22 q_1^0 2q_2^0}
\left\langle P+\frac{\varDelta}{2},n \left| \V \bigg\vert q_1,n_1;q_2,n_2\bigg\rangle\bigg\langle q_1,n_1;q_2,n_2
\bigg\vert \widetilde{\OO}_{1/2}(z)\right| P-\frac{\varDelta}{2},n \right \rangle \,
\frac{1}{P^0+ {\varDelta^0\over2}-q_1^0 - q_2^0},
\label{eq:I_insert:equal:time:quantization}
\end{align}
where $q_{1,2}^0 = \sqrt{\mu_{n_{1,2}}^2 + q_{1,2}^2}$ with $\mu_{n_{1,2}}$ denoting the masses of the $n_{1,2}$-th mesonic states.
Here, we have used the fact that when the unit operator is expanded in the basis of all possible energy eigenstates,
only those intermediate two meson states can yield a nonvanishing contribution:
\beq
 \mathbb{I} = \frac{1}{2!}\sum_{n_1,n_2}\int\frac{d q_1^0 dq_2^0} {\left(2\pi\right)^2 2q_1^0 2 q_2^0} \vert q_1,n_1;q_2,n_2\rangle\langle q_1,n_1;q_2,n_2\vert + \cdots.
\eeq

The first matrix element in the integrand in \eqref{eq:I_insert:equal:time:quantization} can be expressed as
\beq
 \bigg\langle P+{\varDelta\over 2},n \bigg\vert \V \bigg\vert  q_1,n_1; q_2,n_2\bigg\rangle \equiv {2\pi}\delta\left(P^z +  {\varDelta^z \over 2 } - q_1^z - q_2^z \right)
 \widetilde{\varGamma}_{n,n_1,n_2} \left(q_1, P + \varDelta/2 \right).
\label{meson:transition:to:two:meson:matrix:element:ETQ}
\eeq
Analogous to its light-cone counterpart \eqref{meson:transition:to:two:meson:matrix:element}, here we introduce the three-meson vertex function $\widetilde{\varGamma}$
in the equal-time quantization.
It can be expressed as a convolution of three mesonic BGWFs together with the Bogoliubov-chiral angle.
The explicit form of the triple meson vertex function is too lengthy to reproduce here.
We present its full expression in Appendix~\ref{appendix:B}.

For the second matrix element in the integrand in \eqref{eq:I_insert:equal:time:quantization},
it is the $m^\dagger$ component of the
operator $\widetilde\OO_{1/2}(z)$ that renders a nonvanishing result.
Using Bose symmetry and the commutation relation \eqref{Equal:time:m:mdagger:commutation},
one can factorize this matrix element into the product of two clusters:
\begin{align}
&\bigg\langle q,n_1;q_2,n_2\bigg\vert\widetilde{\OO}_{1/2}(z)\bigg\vert P-\frac{\varDelta}{2},n\bigg\rangle
= \bigg\langle q_1, n_1\bigg\vert P-\frac{\varDelta}{2},n\bigg\rangle
 \bigg\langle q_2,n_2\bigg\vert\widetilde{\OO}_{1/2}(z)\bigg\vert \Omega \bigg\rangle + (q_1\leftrightarrow q_2,\; n_1\leftrightarrow n_2).
\label{Reduction:of:O1/2:matrix:element:ETQ}
\end{align}
Apparently, these two terms yield identical contributions to \eqref{eq:I_insert:equal:time:quantization}
after integrating over $q_1$ and $q_2$ and summing over $n_1$ and $n_2$.

The first matrix element in \eqref{Reduction:of:O1/2:matrix:element:ETQ} is simply determined by the orthogonality condition:
\begin{align}
  \bigg\langle q_1,n_1\bigg\vert P-\frac{\varDelta}{2},n\bigg\rangle =  (2\pi) (2q_1^0) \delta_{n, n_1}
  \delta\left(P^z -\frac{\varDelta^z}{2}-q_1^z \right).
\label{eq:state:orth}
\end{align}
Furthermore, the momentum conservation in \eqref{meson:transition:to:two:meson:matrix:element:ETQ} enforces $q_1^z + q_2^z = P^z +  {\varDelta^z \over 2 }$.
By combining these two $\delta$ functions, one can uniquely determine the spatial momenta $q_1^z=P^z-\varDelta^z/2=(1-\xi) P^z$,
$q_2^z=\varDelta^z=2\xi P^z$, as well as $n_1=n$.

As a consequence, one can make the following substitution for energy denominator in \eqref{eq:I_insert:equal:time:quantization}:
\begin{align}
\frac{1}{P^0+{\varDelta^0 \over 2} - q_1^0-q_2^0} \to {1\over \varDelta^0 - q_2^0},
\end{align}
where $q_2^0 \!= \!\sqrt{\mu_{n_2}^2\!+\!q_2^2}\!=\sqrt{\mu_{n_2}^2\!+\!4\xi^2 P^2}$,  and $\varDelta^0 \!=\! \sqrt{(1\!+\!\xi)^2P^2\!+\!\mu_n^2}\!-\!\sqrt{(1\!-\!\xi)^2P^2\!+\!\mu_n^2}\neq q^0$.
Clearly $q^\mu_2$ and $\varDelta^\mu$ are different 2-momenta.

Similar to \eqref{Def:Psi_function:LF:quantization}, we also introduce an auxiliary function $\widetilde{\varPsi}$:
 \begin{align}
 & \widetilde{\varPsi}_{s}\left(x, P, q\right) \equiv \int_{-\infty}^{+\infty}
 \frac{dz}{4\pi} e^{ixP z} \langle q, s\vert \widetilde{\OO}_{1/2}(z) \vert \Omega\rangle.
\label{Def:Psi_function:ET:quantization}
\end{align}
with the single mesonic state labeled by the momentum $q$ and principle quantum number $s$.

After some manipulation on \eqref{eq:GPD_FMF_expnd_V1} , we are able to express $\widetilde{H}^{(1)}_n(x,\xi,P)$
as
\beq
\widetilde{H}_n^{(1)}(x,\xi,P) =  \sum_{r}\left[
        \frac{\widetilde{\varPsi}_r\left(x, P^z, -\varDelta^z\right)}{2q^0\left(-\varDelta^0-q^0\right)}
       \widetilde{\varGamma}_{n,n,r}\left(P_1;P_2\right) +
        \frac{ \widetilde{\varPsi}_r\left(x, P^z, \varDelta^z\right)}{2q^0\left(\varDelta^0-q^0\right)}
        \widetilde{\varGamma}_{n,n,r}\left(P_2;P_1\right)
       \right].
\label{tilde:H1:ETQ:final:expression}
\eeq
where $q_2$ and $n_2$ have been renamed by $q$ and $r$.
The expression of $\widetilde{\varGamma}$ is too lengthy to be reproduced here.
Nevertheless, the interested readers can find its complete expressions in Appendix~\ref{appendix:B}.
The auxiliary function $\widetilde{\varPsi}$
can be worked out and expressed in terms of the BGWGs and Bogoliubov chiral angle:
\begin{align}
 \widetilde{\varPsi}_{r}\left(x, P,q\right) & = \frac{\sqrt{N_c}}{4\pi}\sqrt{2q_r^0}\cos\frac{
       \theta\left(k_1\right)+\theta\left(k_2\right)}{2}\left[
         \varphi_r^+\left(k_1, q\right)
         -\varphi_r^-\left(k_1, q\right)
       \right],
\label{Psi:ETQ:analytical:expression}
\end{align}

{One readily sees that the quasi-GPD $\tilde{H}(x,\xi)$ is an even function of $\xi$.
Making the transformation $x\to -x$  (or equivalently, $k_{1,2} \to - k_{2,1}$) in \eqref{quasi:GPD:tildeH0:expression} and \eqref{tilde:H1:ETQ:final:expression}, using \eqref{eq:BGWF_cpt} and the fact that $\theta(k)$ is an odd function~\cite{Bars:1977ud},
one readily proves that the quasi-GPD $\tilde{H}(x,\xi)$ is an odd function of $x$. Similar to the light-cone GPD,
this trait can be attributed to the charge-conjugation symmetry of a flavor-neutral meson.}

Equations~\eqref{quasi:GPD:tildeH0:expression} and \eqref{tilde:H1:ETQ:final:expression} constitute the major new results of this work,
which express the ${\cal O}(N_c^0)$ quasi-GPD of a flavor-neutral meson in the 't Hooft model in terms of forward- (backward)-moving Bars-Green wave functions
and the Bogoliubov-chiral angle. The support of $x$ is no longer limited in the range $0\le x \le 1$ but becomes unbounded.

\section{Infinite momentum limit of quasi-GPD}
\label{Sec:Infinite:momentum:limit}

The tenet of LaMET is that all the quasi-partonic distributions are expected to converge to the light-cone partonic distributions
when the hadron is boosted to the infinite momentum.
In this section we prove that it is indeed the case for the quasi-GPDs  in the 't Hooft model.

In the infinite momentum limit, only the forward-moving BGWFs survive in \eqref{quasi:GPD:tildeH0:expression},
whereas the backward-moving BGWFs die out. With the aid of \eqref{eq:BGasy} and \eqref{eq:BGasy2},  in the $P^z \to \infty$ limit,
we find asymptotically
\bsq
\begin{alignat}{2}
  & \sin\frac{\theta(k_1)+\theta(k_2)}{2} \to  \Theta(x+\xi)-\Theta(\xi-x), &&\quad  \cos\frac{\theta(k_1)+\theta(k_2)}{2} \to \Theta(\xi-x)\Theta(\xi+x),
\\
  &  \sqrt{E_{1,2}} \varphi^+_n (k_{1,2},P_{1,2}) \to  \sqrt{2\pi} \varphi_n\left(\frac{x\mp\xi}{1\mp \xi}  \right), &&\quad  \sqrt{E_{1,2}} \varphi^+_n (-k_{2,1},P_{1,2}) \to  \sqrt{2\pi} \varphi_n\left(\frac{-x\mp \xi}{1\mp \xi}  \right),
\\
  & \sqrt{2q^0_n} \varphi^+_n (k_1,\pm \varDelta^z) \to \sqrt{4\pi}\varphi_n\left(\pm\frac{x-\xi}{2\xi}\right),
  &&\quad\frac{1}{2q^0(\pm\varDelta^0-q^0)}\to \Theta(\pm \xi)\frac{1}{t-\mu_r^2}.
\end{alignat}
\esq
Substituting these asymptotic expressions into \eqref{quasi:GPD:tildeH0:expression},
one observes that $\widetilde{H}^{(0)}$ indeed reduces to the expression of the light-cone counterpart $H^{(0)}$,
\eqref{eq:LCGPD_exp_cmpct_dglap}.

In the $P^z \to \infty$ limit,  the auxiliary function $ \widetilde{\varPsi}_r(x,P^z,\varDelta^z)$ in \eqref{Psi:ETQ:analytical:expression}
also reduces to its light-cone counterpart $\varPsi_r(x,2\xi)$ in \eqref{Psi:LC:analytical:expression}.
Furthermore, as shown in Appendix~\ref{appendix:A}, in the infinite momentum limit
the three-meson vertex function $\widetilde{\varGamma}_{n,n,r}(P_{1,2},P_{2,1})$ in \eqref{tilde:H1:ETQ:final:expression}
also reduces to its {light-cone counterpart $\varGamma\left(\frac{1\mp\xi}{1\pm \xi},\pm\frac{2\xi}{1\pm\xi}\right)$ in \eqref{eq:H1_IMF_exp}. }
As a consequence, we prove that the $\widetilde{H}^{(1)}$ in \eqref{tilde:H1:ETQ:final:expression}
indeed converges to its light-cone counterpart $H^{(1)}$ in \eqref{eq:LCGPD_exp_cmpct_erbl}.

\section{Forward limit of the light-cone and quasi-GPDs}
\label{sec:foward_limt}

In the forward limit $\xi\to 0$, $t\to 0$, light-cone GPD reduces to the light-cone PDF, and quasi-GPD also reduces to the quasi-PDF.
This relation can be readily seen by setting $\varDelta\to 0$ in the operator definitions of light-cone and quasi-GPD in \eqref{eq:lcgpd_s0} and \eqref{eq:quasigpd_s0}.
Therefore, the purpose of this section is to reinvestigate the light-cone and quasi-PDF, as a byproduct of our studies on light-cone and quasi-GPD.

The advent of LaMET motivated a flurry of studies of quasi-PDFs by employing the two-dimensional QCD as a toy model~\cite{Jia:2018qee,Ji:2018waw,Ma:2021yqx}.
The convergence of quasi-PDFs into the light-cone PDF in the large momentum limit with different meson species has been
numerically verified Ref.~\cite{Jia:2018qee}. Unfortunately, the expression of the quasi-PDF given in \cite{Jia:2018qee} is incomplete,
because the authors of Ref.~\cite{Jia:2018qee} have neglected the contribution from $\tilde{\OO}_{1/2}(z)$ and the next-to-leading Fock component of the mesonic state,
which actually renders a net $\mathcal{O}(N_c^{0})$ contribution.
In the other words, the forward limit of $\tilde{H}^{(1)}$ in \eqref{eq:GPD_FMF_expnd_V1} is absent in Ref.~\cite{Jia:2018qee}.
We take this opportunity to present the complete and correct, expression of the $\mathcal{O}(N_c^{0})$ quasi-PDF in this section.

Taking the forward limit of the quasi-GPD in \eqref{breaking:quasi:GPD:two:pieces},
we can break the quasi-PDF of a flavor-neutral meson into two pieces:
\beq
\tilde{q}_n(x) = \tilde{q}_n^{(0)}(x)+\tilde{q}_n^{(1)}(x),
\label{quasi:PDF:break:into:q0:q1}
\eeq
with
\bseq
\begin{align}
\tilde{q}^{(0)}_n(x,P) =&\int\frac{dz}{2\pi}e^{ixP^z z}{\left\langle P \left| \tilde{\OO}_0(z) \right| P \right \rangle}.
\label{quasi:PDF:q0:def}
\\
\tilde{q}^{(1)}_n(x,P)   =&\int\frac{dz}{2\pi}e^{ixP^zz}\left[ {\left\langle P \left| \tilde{\OO}_{1/2}(z)\frac{1}{P^0-\Hz+i\epsilon}\V \right| P \right \rangle}+{\left\langle P\left| \V\,\frac{1}{P^0-\Hz-i\epsilon}  \tilde{\OO}_{1/2}(z)\right| P\right \rangle}\right],
\label{quasi:PDF:q1:def}
\end{align}
\label{eq:qPDF_FMF_expnd}
\eseq
which are obtained from \eqref{quasi:GPD:two:parts:def} by setting $\varDelta\to 0$.

The analytic expression of $\tilde{q}^{(0)}$ in \eqref{quasi:PDF:q0:def} can be readily obtained from taking the forward limit of $\widetilde{H}^{(0)}$ in \eqref{quasi:GPD:tildeH0:expression}:
\begin{align}
\tilde{q}_n^{(0)}(x,P) =&\frac{P^0}{2 \pi } \sin \theta(xP) \left[\varphi_n^-\left(xP,P\right) \varphi_n^-\left(xP,P\right)+\varphi_n^-\left(-xP,P\right) \varphi_n^-\left(-xP,P\right)\right.
\nn\\
  &\left.+\varphi_n^+\left(xP,P\right) \varphi_n^+\left(xP,P\right)+\varphi_n^+\left(-xP,P\right) \varphi_n^+\left(-xP,P\right)\right],
\label{quasi:PDF:q0}
\end{align}
which agrees with Ref.~\cite{Jia:2018qee}.

$\tilde{q}_n^{(1)}$ represents a new contribution that has been neglected in \cite{Jia:2018qee}.
Taking the forward limit in \eqref{tilde:H1:ETQ:final:expression}, we obtain
\begin{align}
 \tilde{q}_n^{(1)}(x,P) =&  -2\sum_{r}
   \frac{1}{\mu_r^2}\widetilde{\varPsi}_r\left(x, P,0\right)
   \widetilde{\varGamma}_{n,n,r}\left(P;P\right),
\label{quasi:PDF:q1}
\end{align}
with
\beq
  \widetilde{\varPsi}_r\left(x, P,0\right) = \frac{\sqrt{N_c}}{2\pi}\sqrt{2\mu_r}\cos\theta\left(xP\right)\left[
    \varphi_r^+\left(xP,0\right)
    -\varphi_r^-\left(xP,0\right) \right].
\eeq

Equations~\eqref{quasi:PDF:q0} and \eqref{quasi:PDF:q1} represent the complete ${\cal O}(N_c^0)$ expressions of the quasi-PDF of a flavor-neutral meson in the 't Hooft model.
From \eqref{quasi:PDF:q1}, one realizes that one has to include an infinite tower of excited mesonic states to obtain the correct quasi-PDF at leading color,
which is far more involved than naively thought in the preceding work~\cite{Jia:2018qee,Ji:2018waw,Ma:2021yqx}.
In Appendix~\ref{appendix:C}, we make a detailed numerical comparison between the profiles of the complete and correct quasi-PDFs and the incomplete, old ones~\cite{Jia:2018qee},
with different meson momenta and for different meson species.
The effect of the new $\tilde{q}_n^{(1)}(x,P)$ piece becomes less important relative to $\tilde{q}_n^{(0)}(x,P)$ with the increasing quark mass.
However, its effect becomes pronounced if for a light meson carrying soft momentum.

Interestingly, in the infinite momentum limit, the new term $\tilde{q}_n^{(1)}(x,P)$ actually fades away due to $\theta(xP)\to \epsilon(x)\frac{\pi}{2}$, and only the
$\tilde{q}_n^{(0)}(x,P)$ survives, so one reproduces the well-known light-cone PDF of the $n$th excited mesonic state:
\beq
 q_n(x)=\varphi_n(x)^2.
\eeq
In contrast to the light-cone GPD, the light-cone PDF of a meson does not entail the sum over an infinite tower of excited mesonic states.

The vanishing of $\tilde{q}^{(1)}(x,P)$ in the infinite momentum limit is equivalent to the vanishing of the
light-cone GPD $H^{(1)}(x,\xi)$ in the forward limit.
Setting $\xi\to 0$ in \eqref{eq:LCGPD_exp_cmpct_erbl}, one observes that $H^{(1)}(x,0)$ simply vanishes because the integration over $w$
has zero interval. Since the $H^{(1)}(x,\xi)$ GPD is relevant only in the ERBL region ($|x|<\xi$), the support of $x$ shrinks to $x=0$ in the forward
limit. This can be clearly seen in \eqref{Psi:LC:analytical:expression}, where the auxiliary function $\varPsi_r$ becomes nonvanishing only at $x=0$.
Because the GPD is an odd function of $x$, we conclude $H^{(1)}(0,0)=0$.

One can understand this trait by inspecting the definition of the auxiliary function $\varPsi_r$ in \eqref{Def:Psi_function:LF:quantization}.
The forward limit $\xi\to 0$ implies that $q^+\to 0$, where $q^+$ represents the light-cone momentum carried the on-shell mesonic state
in \eqref{Def:Psi_function:LF:quantization}.
It is simply impossible for a meson to carry zero light-cone momentum, so the $\varPsi_r$ function must vanish in the forward limit, and the same is
true for $H^{(1)}$ in the forward limit.

\section{Numerical results of light-cone and quasi-GPDs in 't Hooft model}
\label{sec:num}

In this section, we present a comprehensive numerical study for both the light-cone and quasi-GPDs of a flavor-neutral meson in the 't Hooft model.
We consider the quark GPD of four different species of mesons: chiral pion $\pi_\chi$, physical pion $\pi$, strangeonium $s\bar{s}$ and charmonium $c\bar{c}$.
The quark masses are tuned to reproduce the masses of the lowest-lying states for each meson species in the realistic QCD.
We refer the interested readers to Ref.~\cite{Jia:2017uul} for technical details.
The 't Hooft coupling constant $\lambda={g_s^2N_c}/{4\pi}$ in {\QCDtw} is of mass dimension 2,
which is related to the value of string tension in the realistic QCD by choosing $\sqrt{2\lambda}=340\,\mathrm{MeV}$~\cite{Burkardt:2000uu}.
For simplicity, we will use $\sqrt{2\lambda}$ as the mass unit throughout this section.

\begin{table}[htbp]
   \centering
   \renewcommand{\arraystretch}{1.8}
   \begin{tabular}{cccccccccc}
     Meson  ~~&$m/\sqrt{2\lambda}$  ~~&~~$\mu_0/\sqrt{2\lambda}$~~&~~$\mu_1/\sqrt{2\lambda}$ ~~&$P^z$\\
     \hline
     \hline
     $\pi_\chi$    ~~& 0    ~~&$0$ ~~&$2.43$            &~~$\{2,4,8,16\}\times 0.41\sqrt{2\lambda}$   \\ \hline
     $\pi$    ~~& $0.045$    ~~&$0.41$ ~~&$2.50$             &~~$\{2,4,8,16\}\times \mu_0$       \\\hline
     $s\bar{s}$    ~~&$0.749$   ~~&$2.18$ ~~&$3.72$             &~~$\{1,2,4,8\}\times \mu_0$        \\\hline
     $c\bar{c}$    ~~&$4.19$   ~~&$9.03$ ~~&$10.08$             &~~$\{1/2,1,2\}\times \mu_0$  \\
     \hline
   \end{tabular}
\caption{The values of quark masses, mesons' masses and a variety of meson's momenta (averaged between the initial and final-state mesons).
$\mu_0$ and $\mu_1$ signify the masses of the ground state and the first-excited state.}
\label{Tab:setup}
   \end{table}

For each given quark mass, we calculate the light-cone and quasi-GPDs of both the ground state ($n=0$) meson and the first-excited ($n=1$) state.
We choose three benchmark values for the skewness parameter: $\xi=0.25,\,0.5,\,0.75$.
In Table~\ref{Tab:setup} we enumerate the values of quark masses, mesons' masses, and mesons' momenta $P^z$ used in the calculation.
Note that the momenta carried by the initial and final-state mesons are $(1+\xi)P^z$ and $(1-\xi)P^z$, respectively.

The numerical recipes of solving the 't Hooft equation and Bars-Green equations are elaborated in detail in Ref.~\cite{Jia:2017uul}.
Here we briefly outline some key ingredients.
Following 't Hooft's original method~\cite{tHooft:1974pnl}, we expand the mesonic light-cone wave functions in terms of the following basis functions:
 \beq
 \varphi_n(x)=C_0 x^{\beta_1}(1-x)^{2-\beta_1}+C_1(1-x)^{\beta_2} x^{2-\beta_2}+\sum_{k} C_k\sin(k\pi x).
 \label{eq:lcwfs}
 \eeq
 The parameters $\beta_{1,2}$ are determined by the boundary conditions $\pi \beta_{1,2} \cot(\pi\beta_{1,2})=1-m^2/(2\lambda)$~\cite{tHooft:1974pnl}. In \eqref{eq:lcwfs}, the first two boundary terms determine the asymptotic behavior of LWCF when $x\to 0,\, 1$.

We follow the recipe given in Ref.~\cite{Li:1986gf,Jia:2017uul} to solve the bound-state wave functions in equal-time quantization.
The chiral angle $\theta(p)$ has been solved in very high numerical accuracy in Ref.~\cite{Jia:2017uul}.
The Bars-Green wave functions of a moving meson are expanded in terms of the Hermite polynomials and Gaussian functions~\cite{Li:1986gf,Jia:2017uul}:
 \begin{align}
\varphi_n^{\pm}(xP,P)=\sum_{k} C^{n,\pm}_k H_k\left(\frac{\alpha P}{2}(1-2x)\right) e^{-\frac{\alpha^2 P^2(1-2x)^2}{8}}.
 \end{align}
$\alpha$ is a variational parameter that is tuned to minimize the mass of the ground state.

A sum over an infinite tower of excited mesonic states is encountered in evaluating the $H^{(1)}$ component of the light-cone GPD in \eqref{eq:LCGPD_exp_cmpct_erbl},
as well as the $\widetilde{H}^{(1)}$ piece of the quasi-GPD in \eqref{tilde:H1:ETQ:final:expression}. We truncate the sum with $r<N_{\rm max}$.
With trial and error, we find that for $N_{\rm max}=48$, the sum has already exhibits satisfactory convergence behavior.

\begin{figure}[htbp]
   \includegraphics[width=0.9\linewidth]{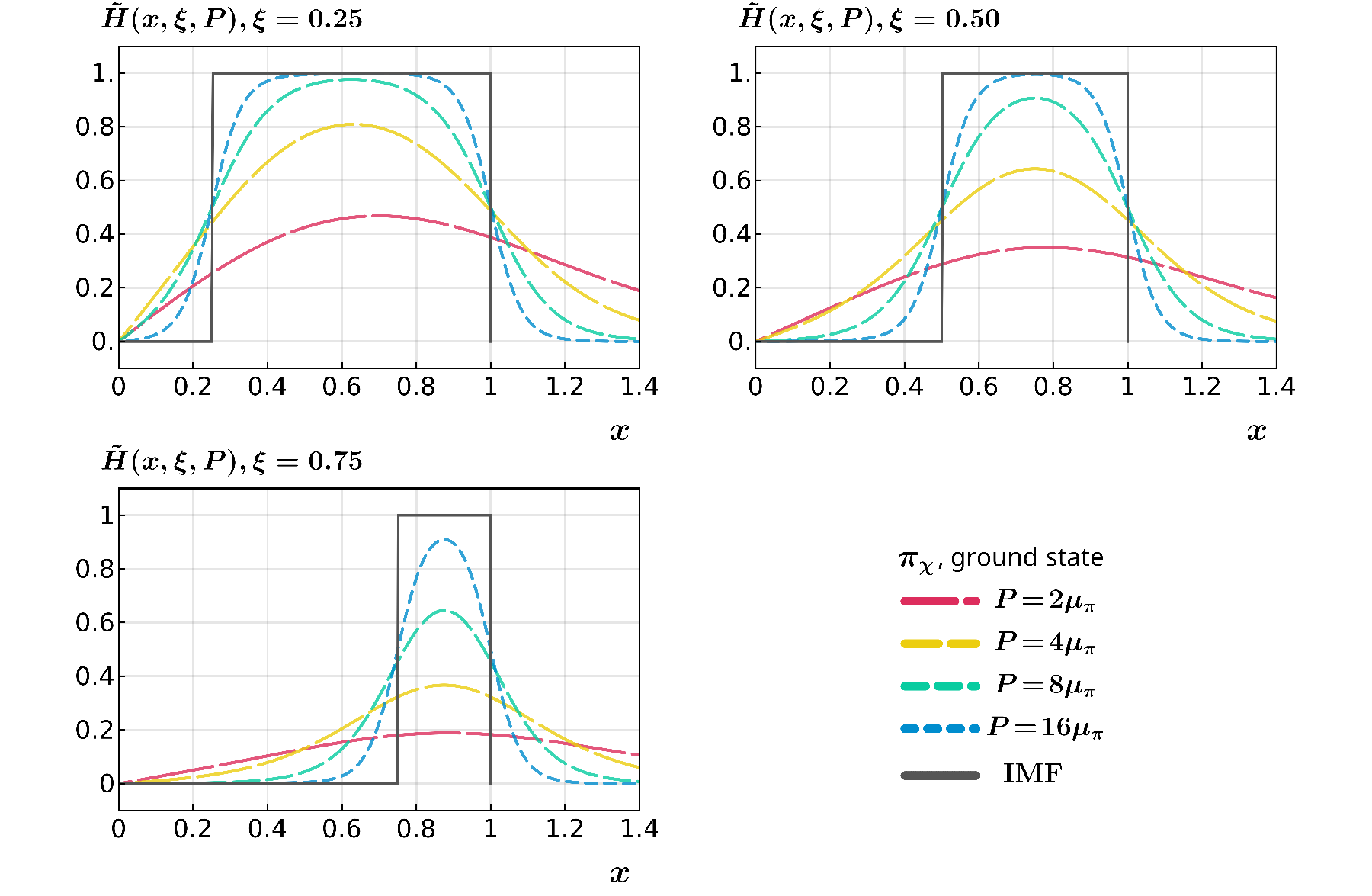}
   \caption{Light-cone and quasi-GPDs of the ground state ($n=0$) chiral pion. $\mu_\pi$ stands for the mass of ground state physical pion.}
   \label{fig:chipiGPD_n0}
 \end{figure}

 \begin{figure}[htbp]
   \includegraphics[width=0.9\linewidth]{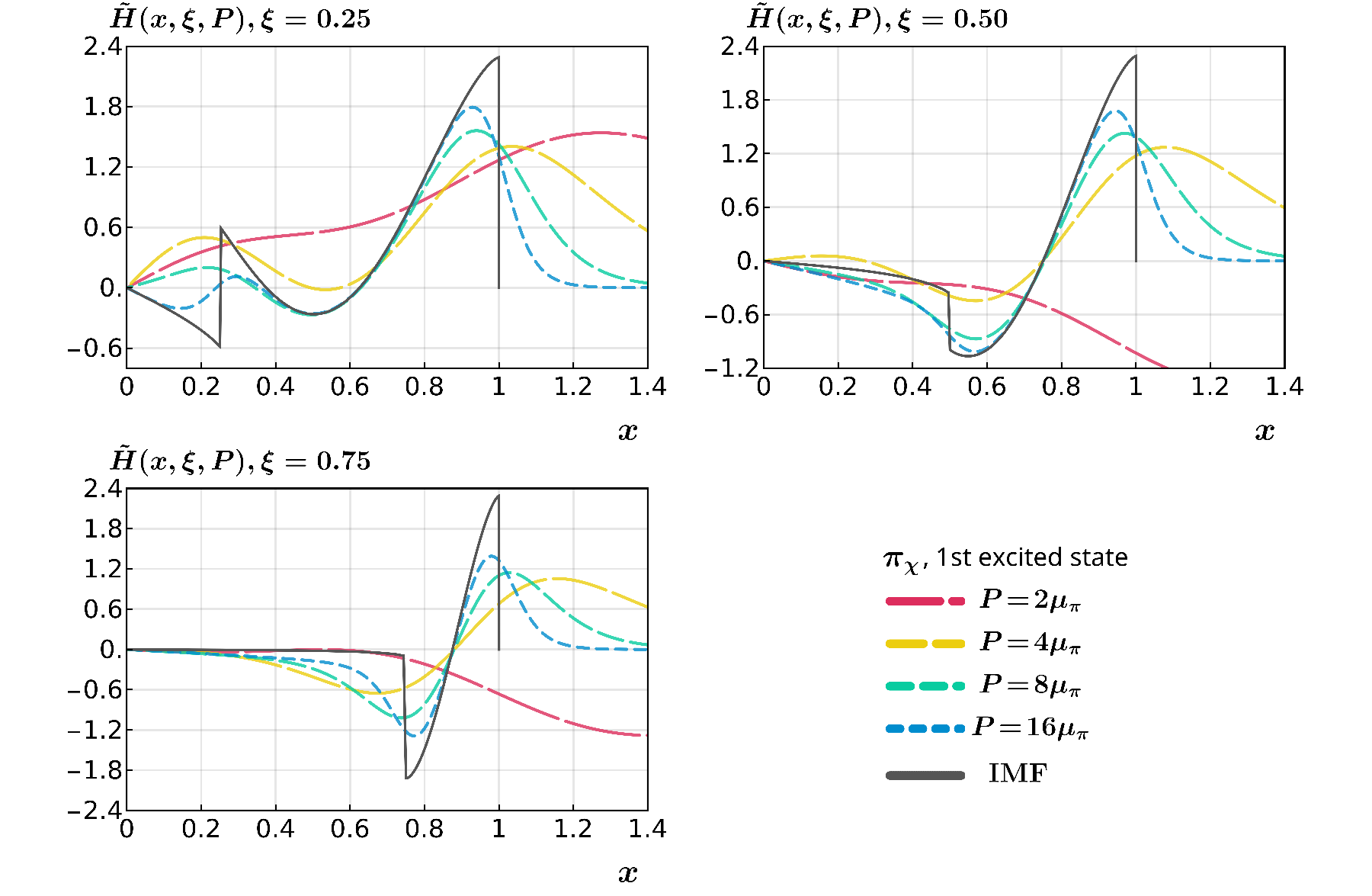}
   \caption{Light-cone and quasi-GPDs of the first-excited ($n=1$) state in the chiral pion family.}
   \label{fig:chipiGPD_n1}
 \end{figure}

 \begin{figure}[htbp]
   \includegraphics[width=0.9\linewidth]{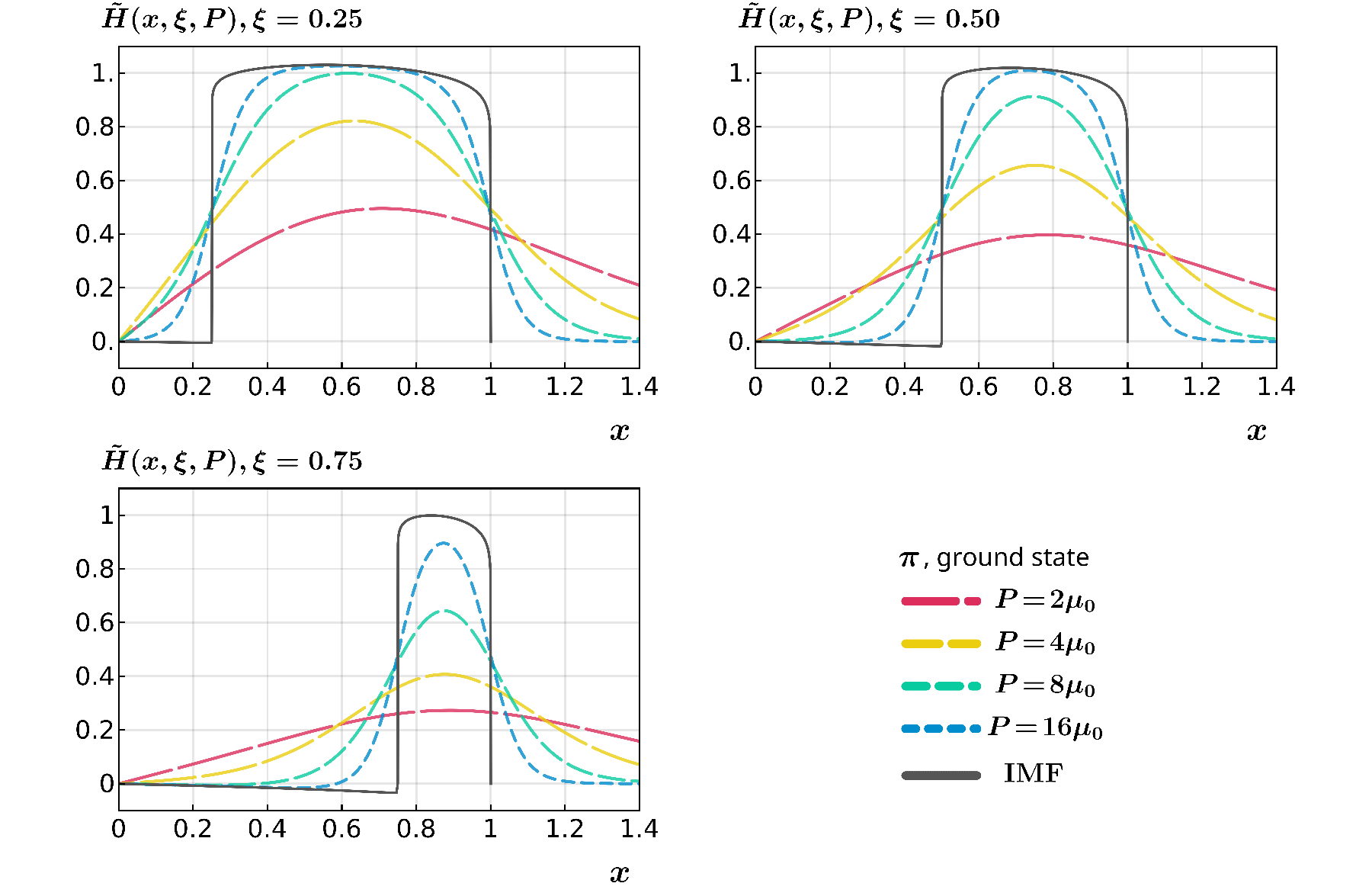}
   \caption{Light-cone and quasi-GPDs of the ground state physical pion. }
 \label{fig:piGPD_n0}
 \end{figure}

 \begin{figure}[htbp]
   \includegraphics[width=0.9\linewidth]{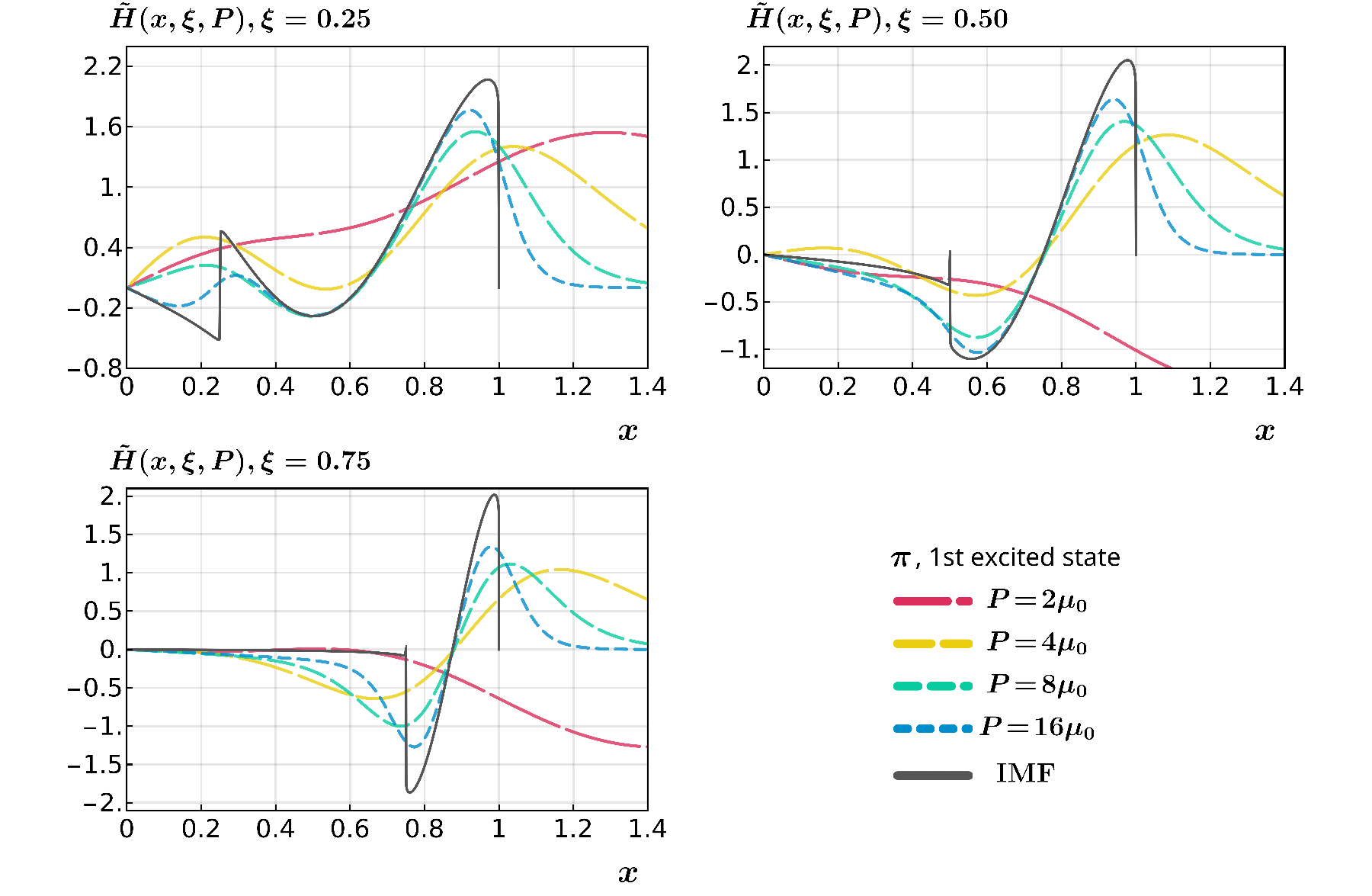}
   \caption{Light-cone and quasi-GPDs of the first-excited pionic state. }
   \label{fig:piGPD_n1}
 \end{figure}

 \begin{figure}[htbp]
   \includegraphics[width=0.9\linewidth]{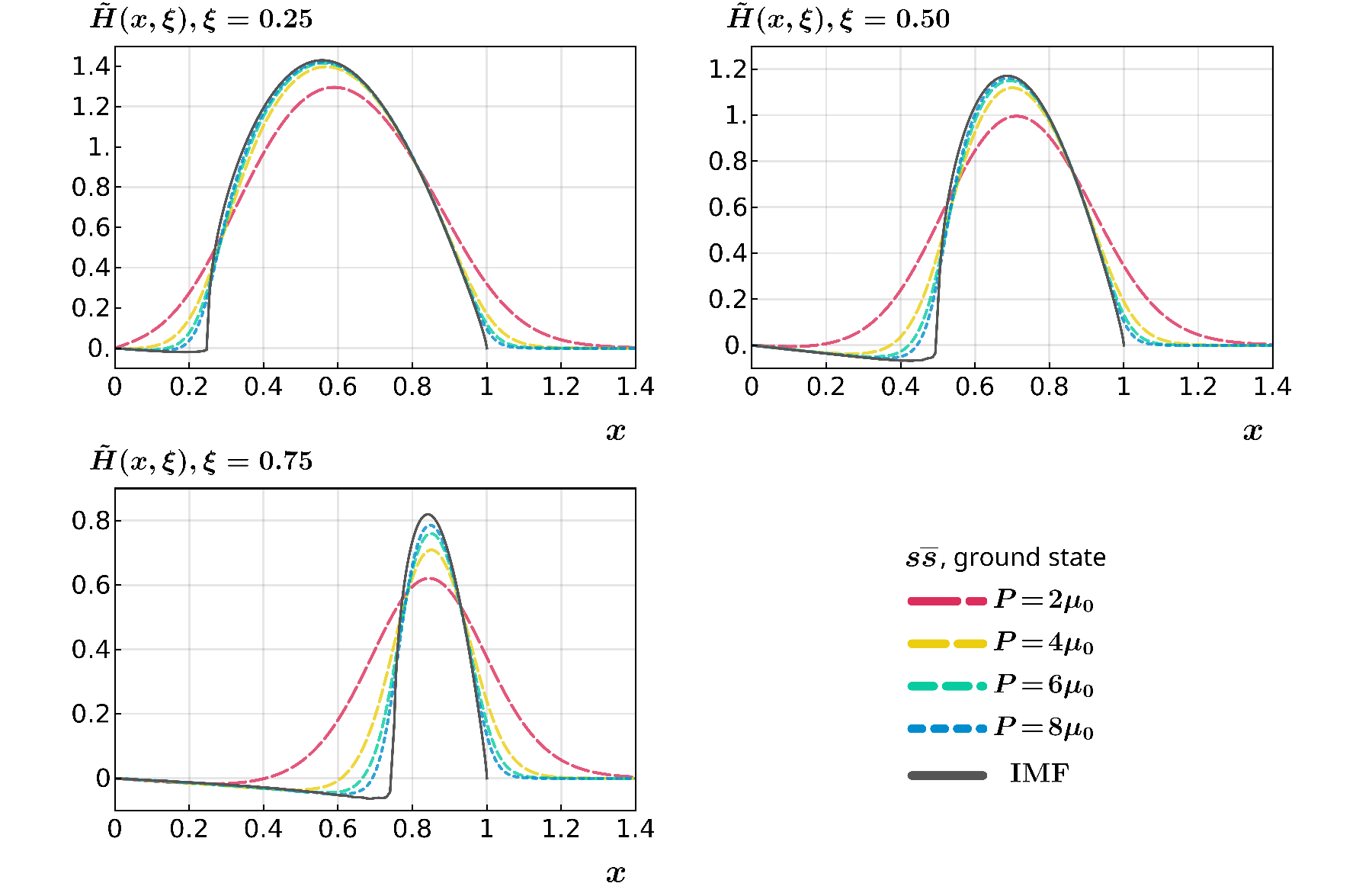}
   \caption{Light-cone and quasi-GPDs of the ground state strangeonium.}
   \label{fig:ssGPD_n0}
   \end{figure}

   \begin{figure}[htbp]
     \includegraphics[width=0.9\linewidth]{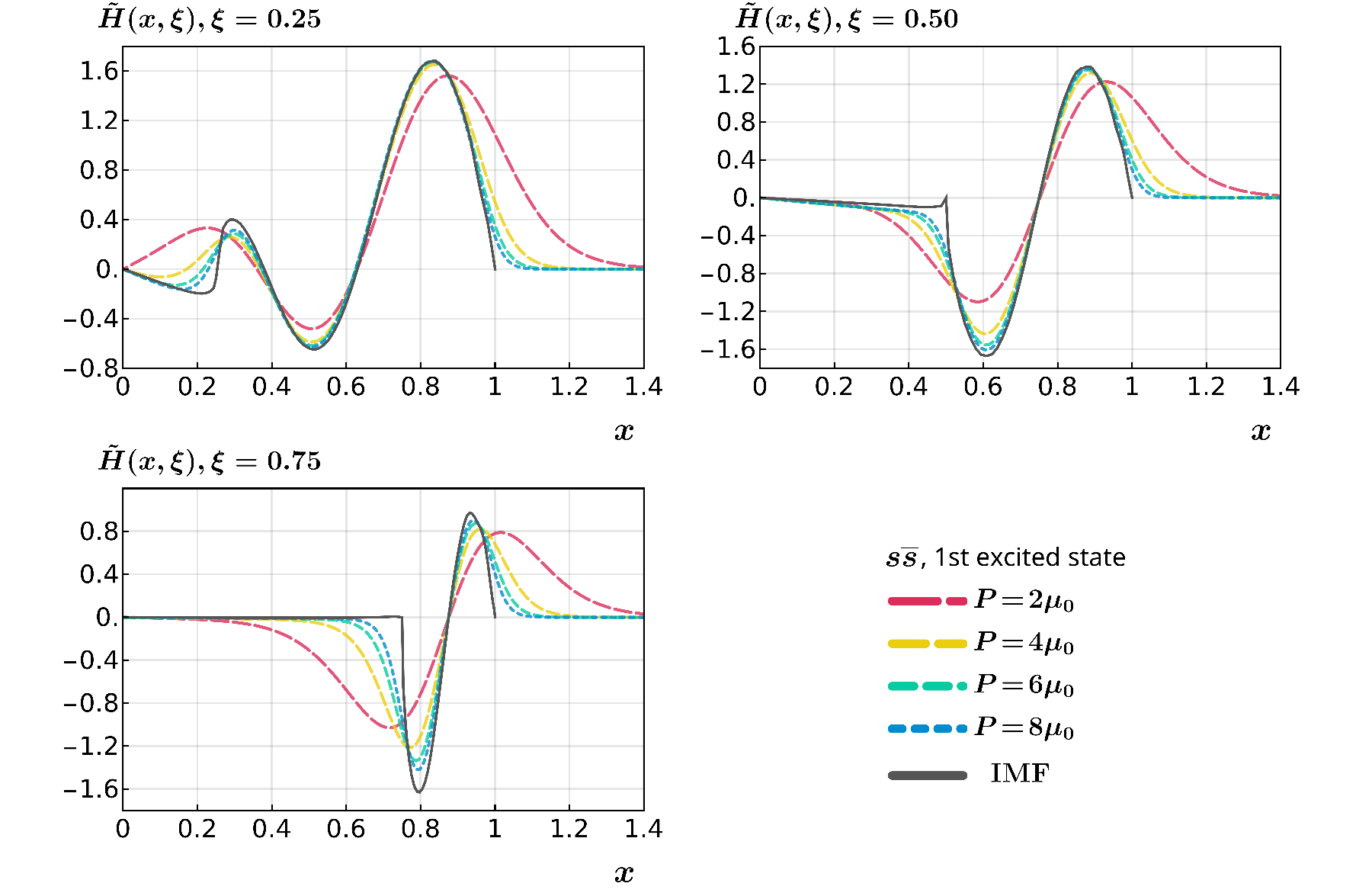}
     \caption{Light-cone and quasi-GPDs of the first-excited strangeonium state.}
     \label{fig:ssGPD_n1}
     \end{figure}

 \begin{figure}[htbp]
   \includegraphics[width=0.9\linewidth]{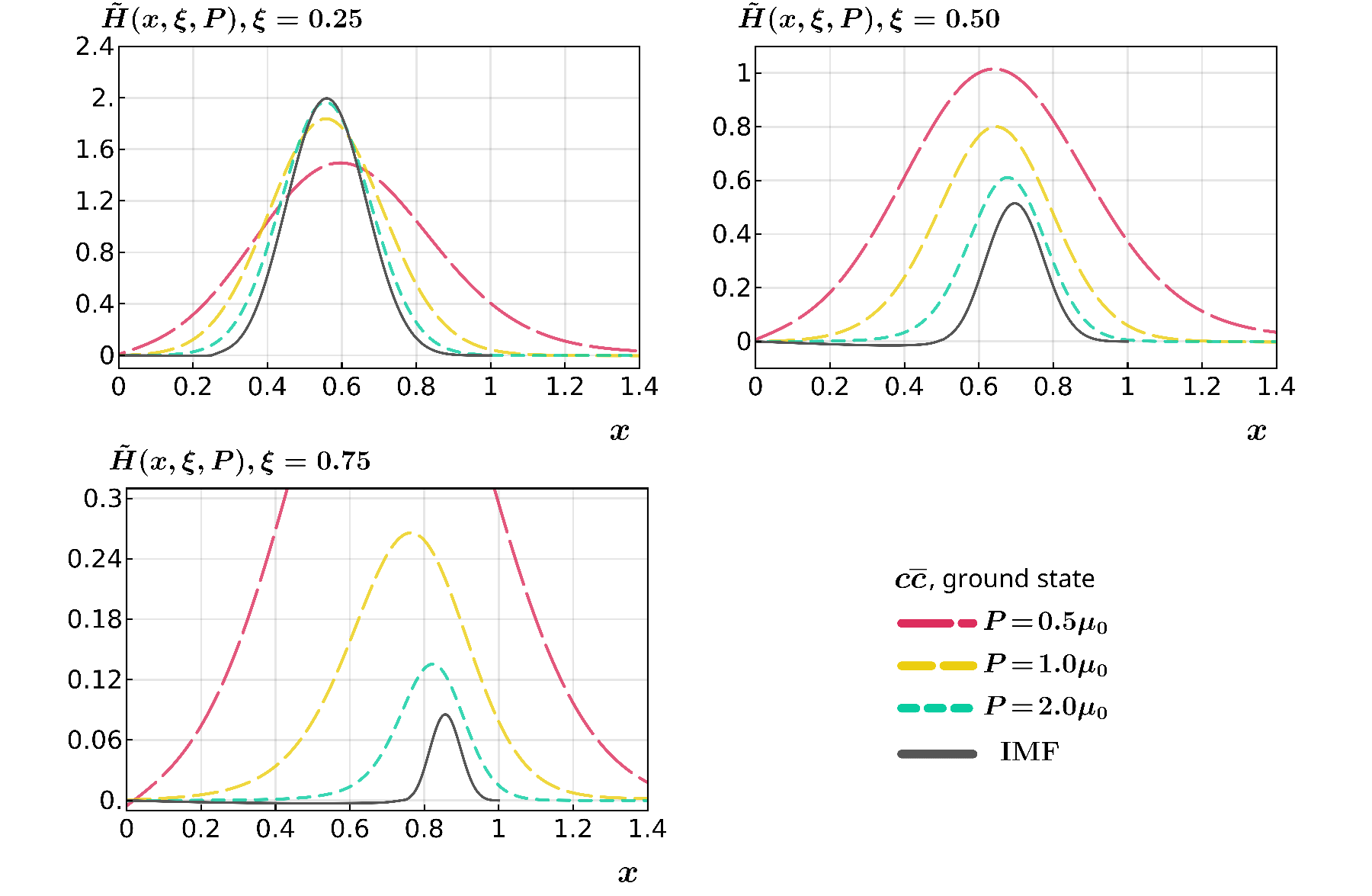}
   \caption{Light-cone and quasi-GPDs of the ground state charmonium. }
   \label{fig:ccGPD_n0}
 \end{figure}

 \begin{figure}[htbp]
   \includegraphics[width=0.9\linewidth]{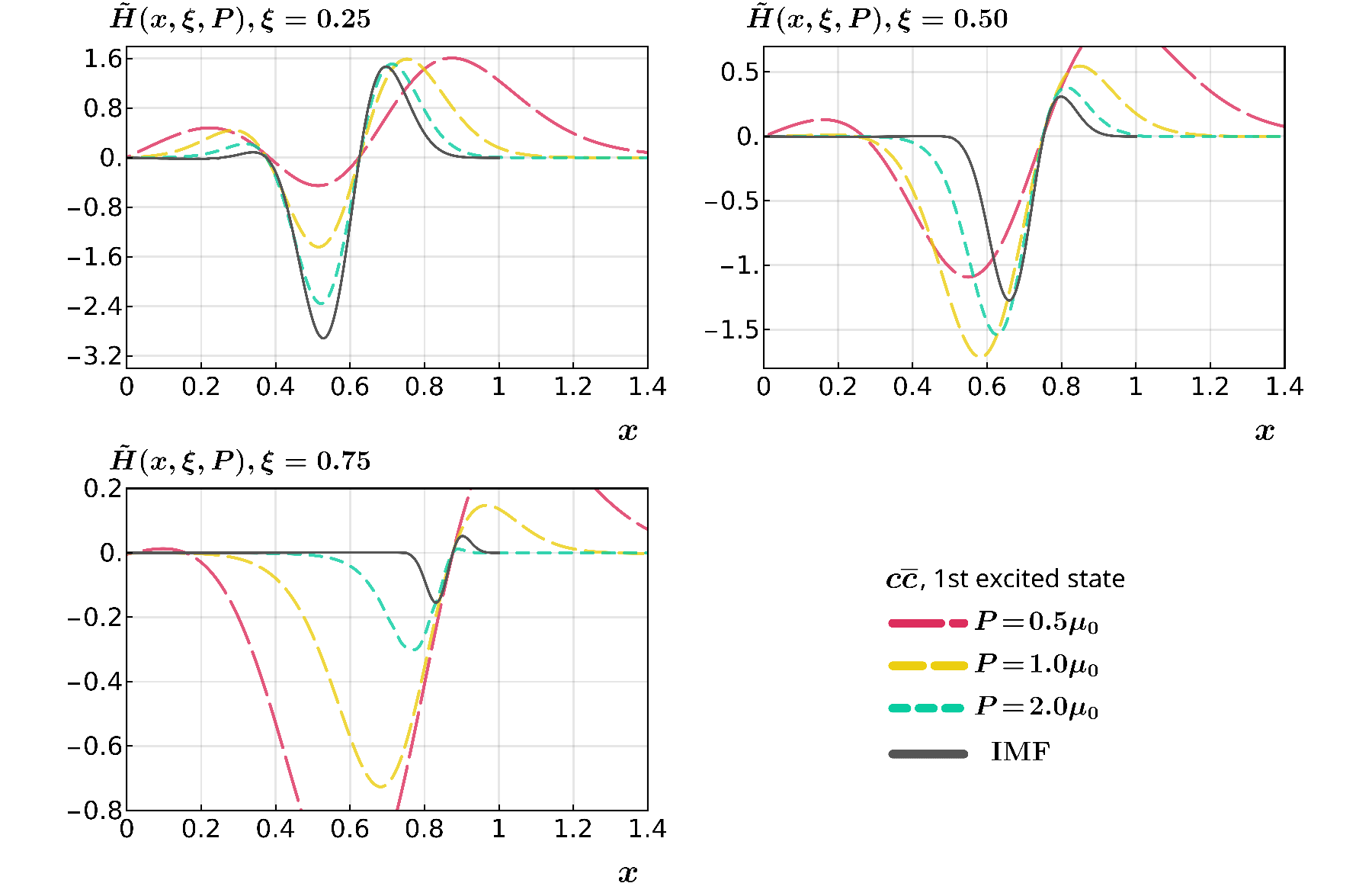}
   \caption{Light-cone and quasi-GPDs of the first-excited charmonium state. }\label{fig:ccGPD_n1}
 \end{figure}

The profiles of the light-cone and quasi-GPD for the $n=0$ and $n=1$ mesons with four different quark masses are plotted in Figs.~\ref{fig:chipiGPD_n0}-\ref{fig:ccGPD_n1}.
For simplicity we have only plotted he $x>0$ region. The plots can be straightforwardly extended to negative $x$ regime
since the light-cone and quasi-GPDs are odd functions of $x$.

From Fig.~\ref{fig:chipiGPD_n0} and Fig.~\ref{fig:piGPD_n0}, we observe an interesting pattern of the GPDs of the ground states $\pi_\chi$ and $\pi$,
that the light-cone GPD of ($\pi$) $\pi_\chi$ (nearly) vanishes in the ERBL region $|x|<\xi$.
Recall that the LCWF of the massless chiral pion (with vanishing quark mass) is just the product of two Heaviside step functions $\varphi_0(x)=\Theta(x)\Theta(1-x)$. Substituting this expression into \eqref{eq:LCGPD_exp_cmpct_erbl},
one can prove that the combination $\varphi_0\left(\frac{2\xi-w}{1+\xi}\right)-\varphi_0\left(\frac{2\xi+y}{1+\xi}\right)$ inside the square bracket of
\eqref{eq:LCGPD_exp_cmpct_erbl} strictly vanishes; therefore, the $H^{(1)}$ exactly vanishes in the ERBL region.
The shape of LCWF of a physical pion (with $m=0.045\sqrt{2\lambda}$, $\beta_1=0.024$, and $\beta_2=1.976$) is quite close to that of the chiral pion, which has a wide plateau in the middle and very steep rises and fall near the end points $x=0,\:1$~\cite{Jia:2017uul}.
Therefore, we also observe a near vanishing of its light-cone GPD in the ERBL region.

From Figs.~\ref{fig:chipiGPD_n0}-\ref{fig:ccGPD_n1}, one observes a clear pattern for all different types of mesons:
with the increasing of the momentum, the quasi-GPDs in the physical support $-\xi<x<1$ tend to approach the light-cone GPD,
while the quasi-GPDs in the unphysical support $x<-\xi$ and $x>1$ fade away. This finding provides numerical verification of the
analytical proof in Sec.~\ref{Sec:Infinite:momentum:limit}: the quasi-GPD is expected to converge to its light-cone counterpart in the
infinite momentum limit. Therefore, our numerical results can be viewed as the support for the validity of the LaMET in two-dimensional QCD.

For a given average meson momentum $P$, by varying the values of the skewness parameter, we observe from Figs.~\ref{fig:chipiGPD_n0}-\ref{fig:ccGPD_n1} a notable pattern:
the speed for the quasi-GPD to approach the light-cone GPD with larger $\xi$ is slower than with smaller $\xi$.
This pattern may be partly attributed to a simple kinematic effect. The momentum carried by the initial-state meson is $(1-\xi)P^z$.
Even if the averaged momentum $P^z$ is large enough, the $1-\xi$ factor suppresses the initial-state's momentum, especially for the large $\xi$,
thus the pace of approaching the light-cone GPD from the quasi-GPD gets slowed down with increasing $\xi$.
This problem may pose an obstacle for the attempts to calculate GPD with wide kinematic coverage in the LaMET approach, especially when $\xi$ becomes close to 1.

\begin{figure}[htb]
\centering
\includegraphics[width=0.9\linewidth]{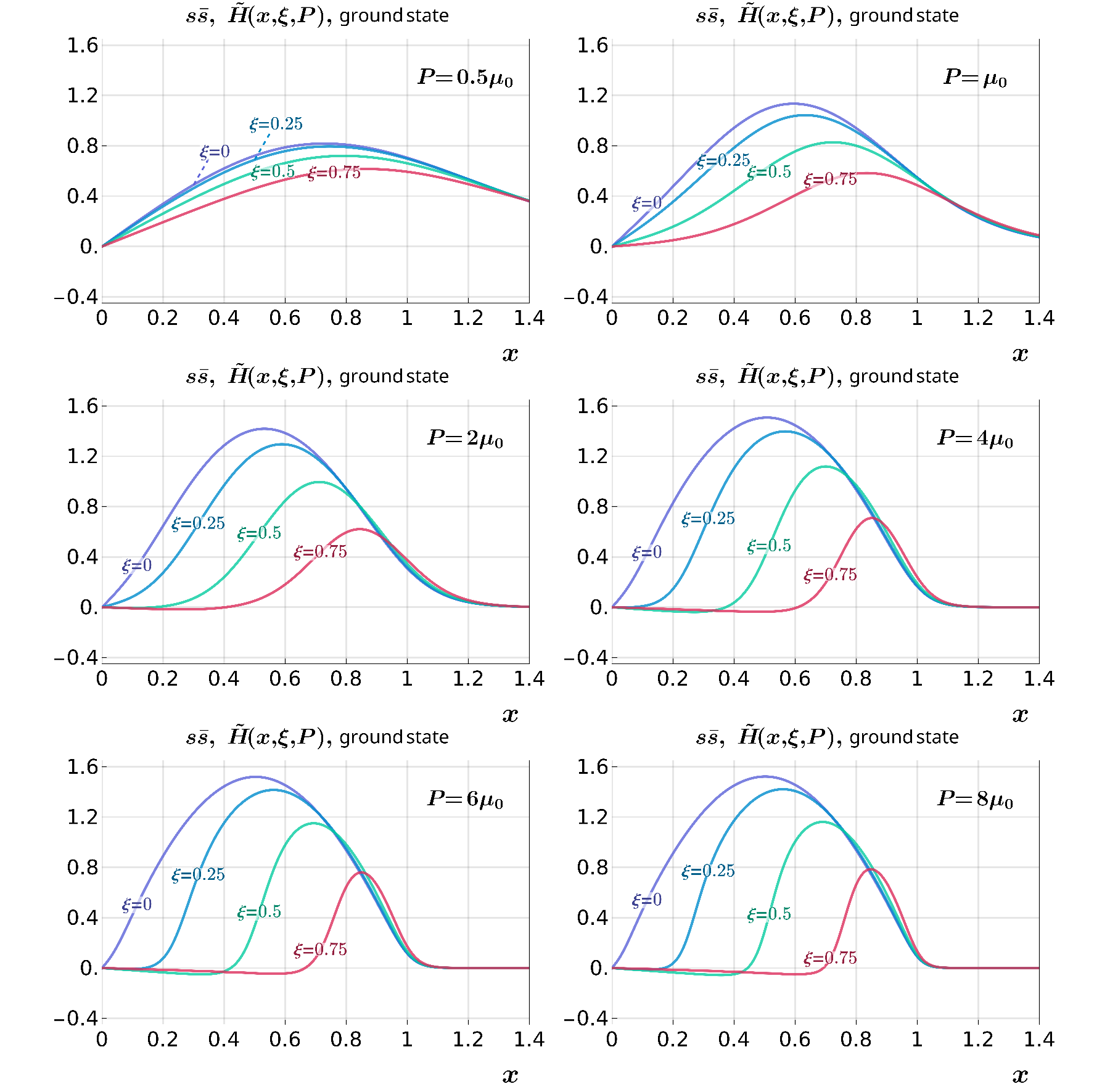}
\caption{quasi-GPD of the ground state strangeonium. In each plot, we juxtapose various quasi-GPDs with the same average momentum $P$ yet with different values of skewness.}
\label{fig:skwdepn0}
\end{figure}

\begin{figure}[htb]
\centering
\includegraphics[width=0.9\linewidth]{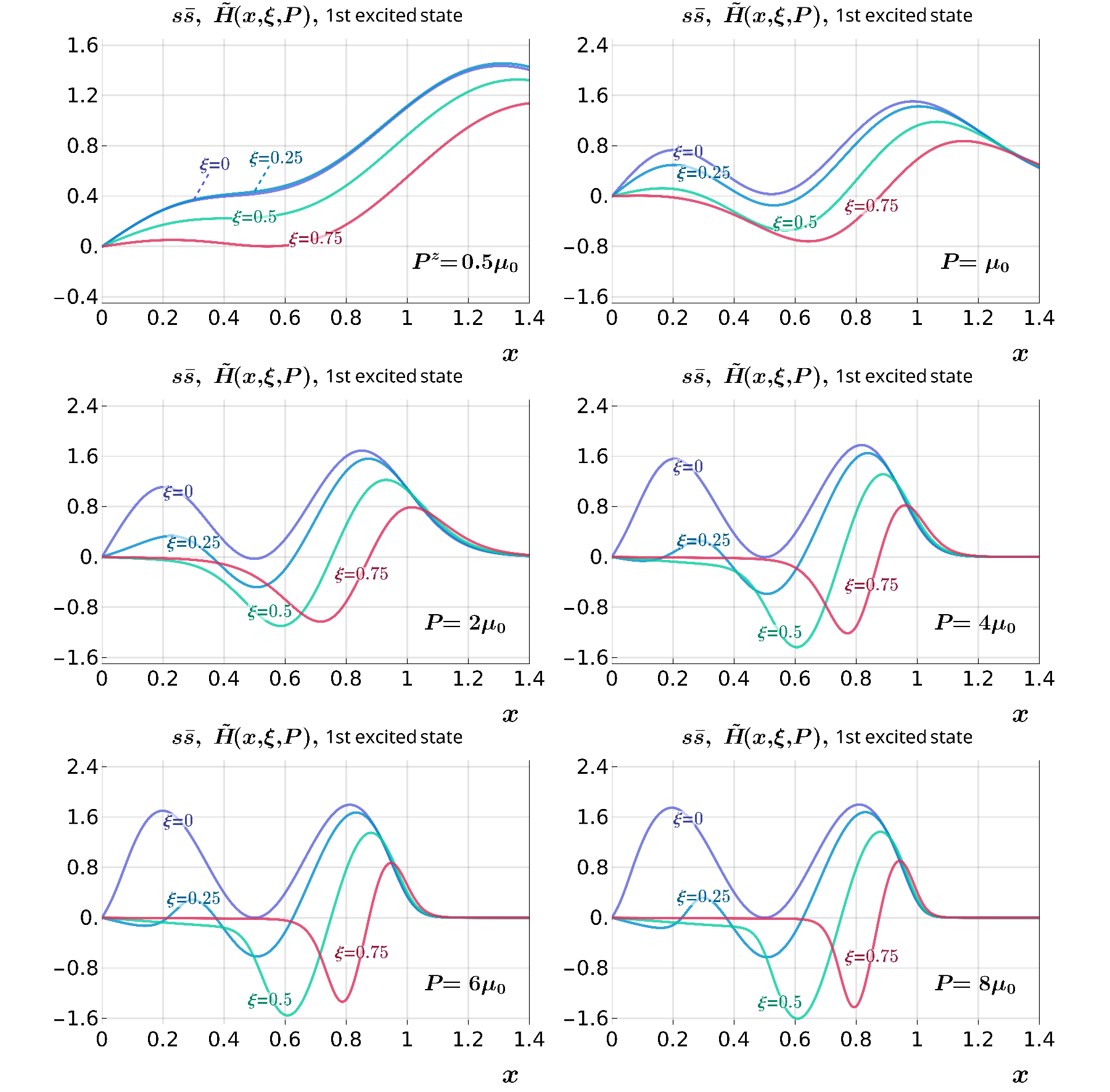}
\caption{Same as Fig.~\ref{fig:skwdepn0}, but the meson is replaced with the first-excited strangeonium state.}
\label{fig:skwdepn1}
\end{figure}

It is also interesting to examine how the dependence of the quasi-GPDs on the momentum transfer $\varDelta$ evolves with meson momentum.
In Figs.~\ref{fig:skwdepn0} and~\ref{fig:skwdepn1}, we juxtapose the quasi-GPDs of the lowest-lying and first-excited strangeonia states
with the same average meson momentum $P^z$ yet with different skewness (recall $\xi$ and $\varDelta$ are interrelated in {\QCDtw}).
We choose several different values of $P^z$, ranging from half of to eight times the strangeonium mass.
We observe that disparity among the quasi-GPDs with different values of $\xi$ become amplified with the increasing mesonic momentum $P^z$.
Nevertheless, as $P^z>4\mu_0$, the disparity pattern appears to become frozen, indicating the quasi-GPD with each $\xi$ has approached the respective IMF limit.

\begin{figure}[htb]
\centering
\includegraphics[width=0.9\linewidth]{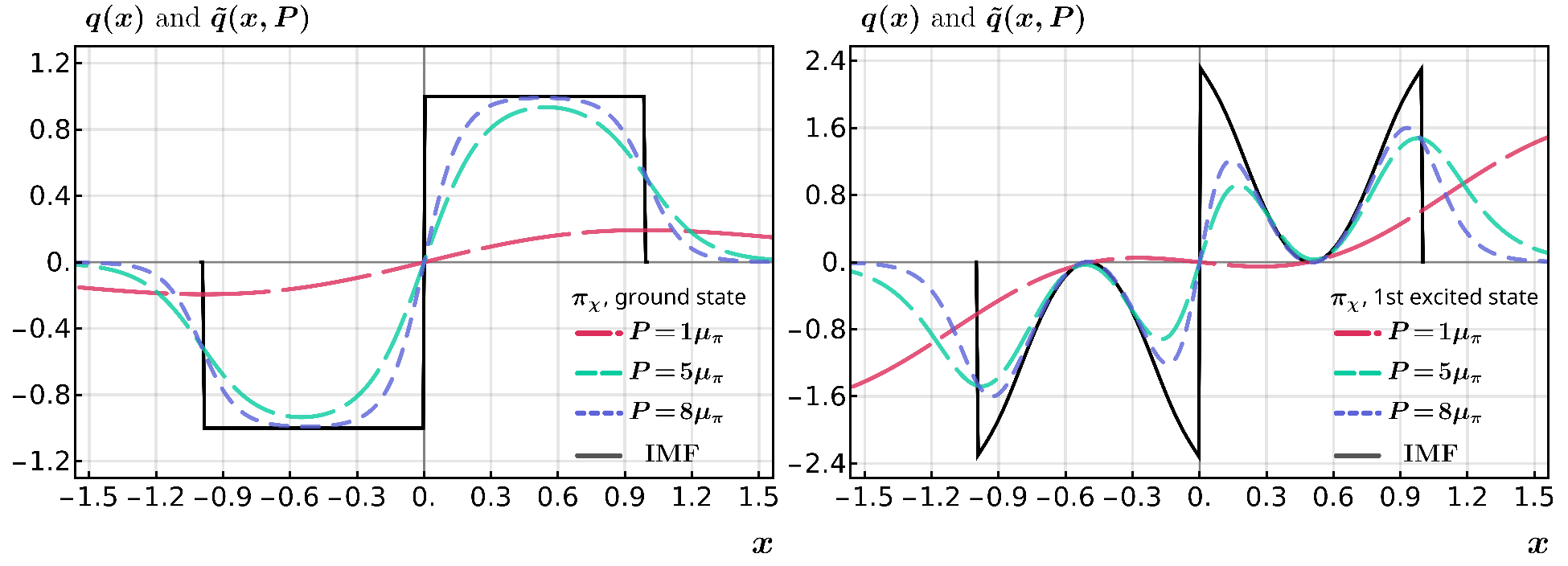}
\includegraphics[width=0.9\linewidth]{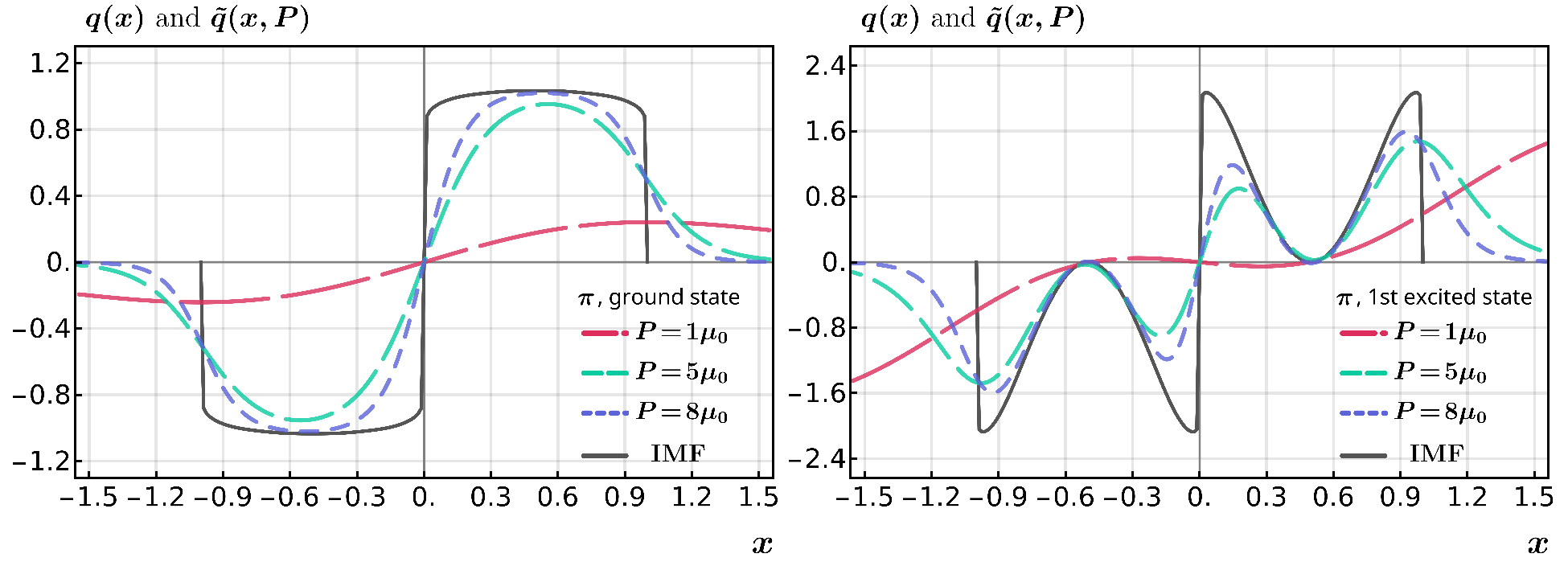}
\caption{Numerical results for light-cone and quasi-PDFs of light mesons ($\pi_\chi$ and $\pi$). The solid curves represent the quasi-PDFs
obtained from \eqref{quasi:PDF:break:into:q0:q1}, by taking the forward limit of the quasi-GPD.
}\label{fig:pdfrev_light}
\end{figure}

\begin{figure}[htb]
\centering
\includegraphics[width=0.9\linewidth]{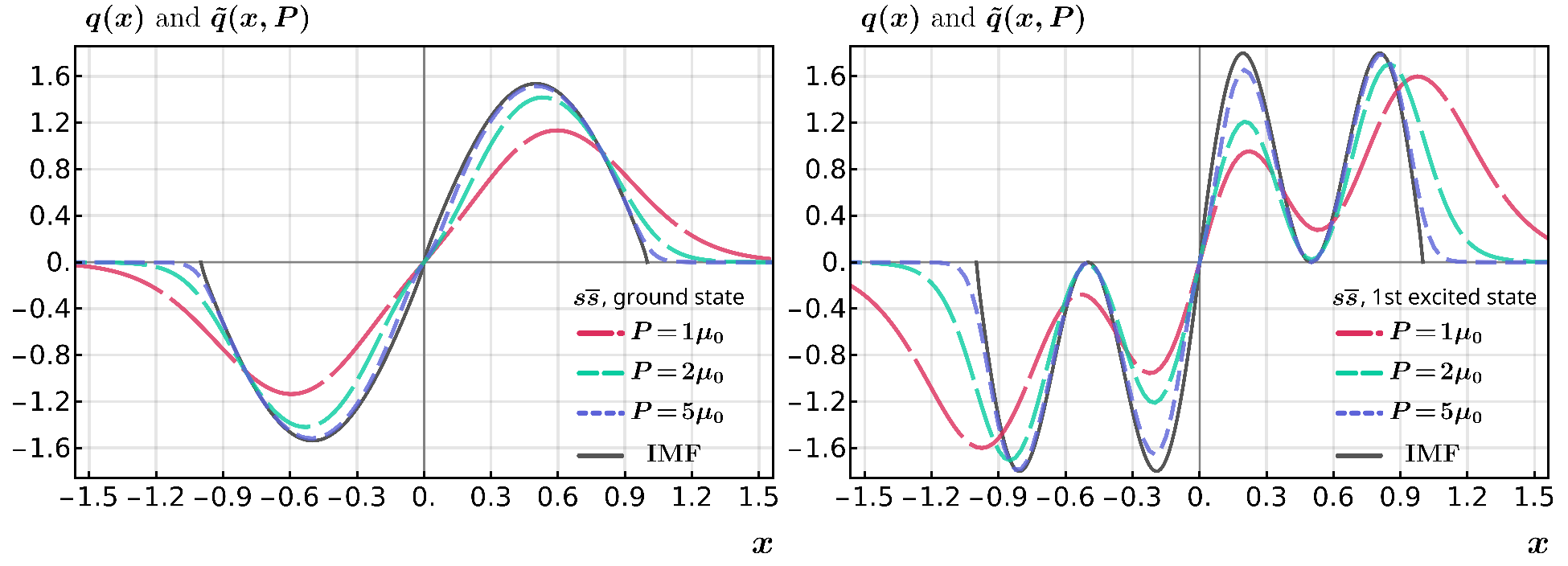}
\includegraphics[width=0.9\linewidth]{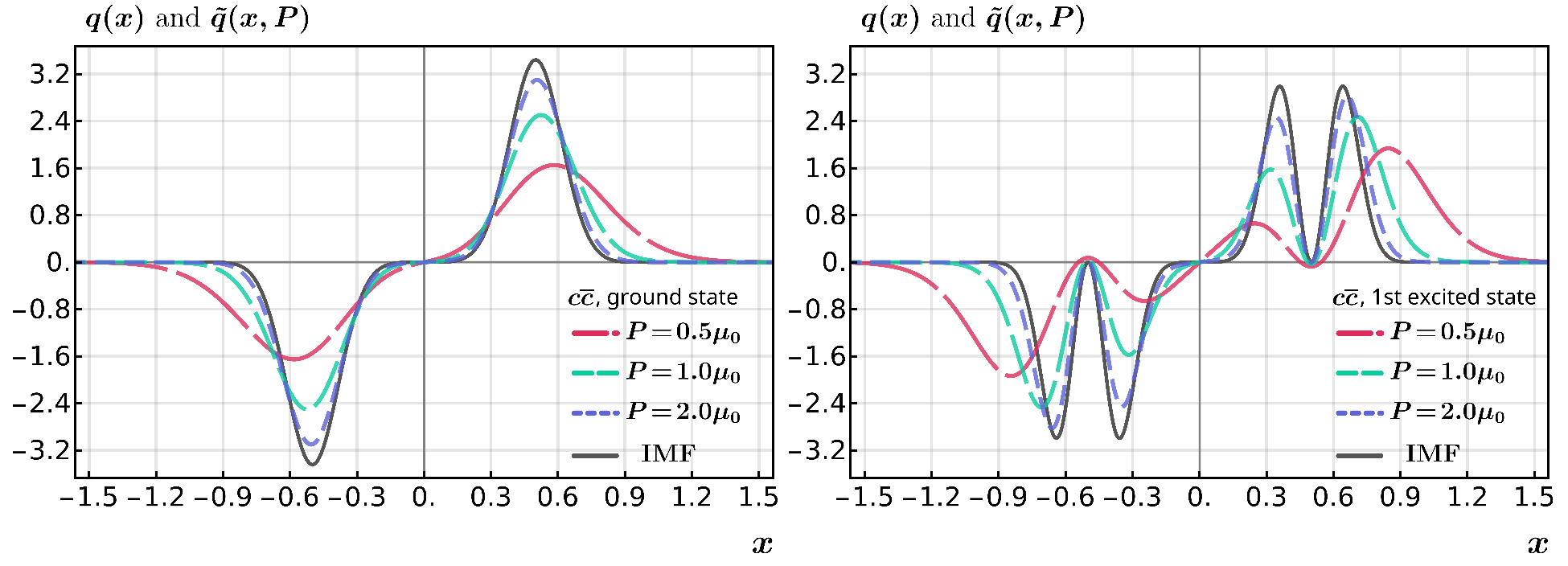}
\caption{Same as Fig.~\ref{fig:pdfrev_light}, but with mesons replaced by the strangeonium and charmonium.}
\label{fig:pdfrev_heavy}
\end{figure}

In Sec.~\ref{sec:foward_limt} we obtain the expressions of the quasi-PDFs by taking the the forward limit $\varDelta\to 0$ from the quasi-GPDs.
As indicated in \eqref{quasi:PDF:break:into:q0:q1}, the complete ${\cal O}(N_c^0)$ quasi-PDF can be decomposed into
$\tilde{q}_n(x,P) = \tilde{q}_n^{(0)}(x,P)+\tilde{q}_n^{(1)}(x,P)$, where
the $\tilde{q}_n^{(1)}(x,P)$ piece arising from the higher Fock state has been missed in the preceding work~\cite{Jia:2018qee,Ji:2018waw,Ma:2021yqx}.
In Figs.~\ref{fig:pdfrev_light} and~\ref{fig:pdfrev_heavy} we plot the quasi and light PDFs for different types of mesons,
starting from the correct expression \eqref{quasi:PDF:break:into:q0:q1} for quasi-PDFs.

It turns out that the only case where the effect of the new $\tilde{q}^{(1)}(x,P)$ piece becomes pronounced is for a soft light hadron ($\pi_\chi$ or $\pi$).
When the meson is boosted to large momentum, the contribution of $\tilde{q}^{(1)}(x,P)$ quickly dies out.
We devote Appendix~\ref{appendix:C} to a detailed numerical comparison between the magnitudes of the $\tilde{q}^{(0)}(x,P)$ and $\tilde{q}^{(1)}(x,P)$.

\section{summary}
\label{sec:sum}

As the generalization of collinear PDF to off-forward kinematics,
GPD entails rich information on the multidimensional structure of nucleon.
The determination of GPDs from the first principle of QCD is highly desirable.
The LaMET approach provides a promising program that allows one to directly extract the $x$ dependence of the nucleon GPD with nonzero skewness
by comporting the quasi-GPDs in the lattice. However, because of greater theoretical complexity and the expensive computational cost,
the exploration along this direction is still in the beginning phase.

On the other hand, it might be rewarding to glean some lessons about the light-cone and quasi-GPDs from toy models of QCD.
Among several solvable field theory models, the 't Hooft model, {\it i.e.}, (\QCDtw) in the $N_c\to\infty$ limit,
occupies a special position because it resembles the realistic ${\rm QCD}_4$ in several aspects,
such as color confinement, the Regge trajectory, the nonzero quark condensate and (naive) asymptotic freedom.
In this work, we have conducted a thorough investigation of the light-cone and quasi-GPDs of a flavor-neutral meson
in the 't Hooft model. We hope that our study can shed some light on our understanding of the quasi and light-cone GPDs.

Employing the Hamiltonian approach and bosonization procedure,  we deduce the functional form of the
light-cone GPD, expressed in terms of the meson's light-cone wave functions in the framework of light-front quantization.
We also derive the functional form of the quasi-GPD,  expressed in terms of the meson's Bars-Green wave functions
together with the Bogoliubov-chiral angle in the framework of equal-time quantization.
We have verified the key assumption of the LaMET for the off-forward parton distributions in two-dimensional QCD,
both analytically and numerically, that the quasi-GPDs do approach their light-cone counterparts
when the meson is boosted to the infinite momentum limit.
We find that the quasi-GPD with small skewness parameter tends to converge to the light-cone GPD faster than that with large $\xi$.
This pattern may be partly ascribed to the $1-\xi$ suppression factor affiliated
with the momentum carried by the initial-state meson.

Taking the forward limit of the quasi-GPD, we also obtain the analytical expression of the quasi-PDF as a byproduct.
We find that our preceding work~\cite{Jia:2018qee} conveys an incomplete expression for quasi-PDF,
which misses a piece of leading color [${\cal O}(N_c^0)$] contribution stemming from the higher-order Fock component.
We take this opportunity to correct the mistake made in Ref.~\cite{Jia:2018qee} and present a complete expression for quasi-PDFs.
This new contribution to the quasi-PDF quickly fades away when the hadron gets heavier or boosted with larger momentum,
and it has an significant impact only for soft light hadrons.

\begin{acknowledgments}
We are grateful to Jichen Pan for participating in the early stage of this work.
We acknowledge the High Performance Computing Center of Central South University for supporting numerical computation in this work.
The work of Y. J. and Z.-W. M. is supported in part by the National Natural Science
Foundation of China under Grants No. 11925506, No. 12070131001 (CRC110 by DFG and NSFC).
The work of X.-N.~X. is supported by the National Natural Science Foundation of China under Grant No.~12275364.
\end{acknowledgments}

\appendix

\section{Interacting Hamiltonian and three-meson vertex in light-front quantization}
\label{appendix:A}

In Sec.~\ref{H:approach:LF:quantization}, we decompose the light-front Hamiltonian  $\mathbb{H}_{\mathrm{LF}}$ in three pieces,
as indicated in \eqref{eq:HLF_full}. There is an piece representing the mesonic interacting Hamiltonian, $\Vlf$,
starting with ${\cal O}(1/\sqrt{N_c})$. We have listed one typical term in \eqref{V:LF:one:typical:example}.
Here we present the complete expression of the $\Vlf$:
\begin{align}
  \notag \Vlf=&\frac{\lambda}{(2\pi)^{3/2}\sqrt{N_c}}\sum_{n_1,n_2,n_3}\int_0^{\infty} dk_1^+dk_2^+dk_3^+dk_4^+dk_5^+\Bigg[\frac{-\delta \left(k_1^++k_2^+-k_3^++k_4^+\right) }{\left(k_2^++k_4^+\right)^2\sqrt{{\left(k_1^++k_2^+\right) \left(k_3^++k_5^+\right) \left(k_4^++k_5^+\right)}}}
\\
   &\times  \varphi_{n_1}\left(\frac{k_1^+}{k_1^++k_2^+}\right) \varphi_{n_2}\left(\frac{k_3^+}{k_3^++k_5^+}\right) \varphi_{n_3}\left(\frac{k_4^+}{k_4^++k_5^+}\right) m_{n_1}\left(k_1^++k_2^+\right)m_{n_2}^{\dagger }\left(k_3^++k_5^+\right)m_{n_3}\left(k_4^++k_5^+\right) \notag \\
   \notag  &+\frac{-\delta \left(k_1^++k_2^++k_3^+-k_4^+\right) }{\left(k_1^++k_3^+\right)^2\sqrt{{\left(k_1^++k_2^+\right) \left(k_3^++k_5^+\right) \left(k_4^++k_5^+\right)}} }\varphi_{n_1}\left(\frac{k_2^+}{k_1^++k_2^+}\right) \varphi_{n_2}\left(\frac{k_3^+}{k_3^++k_5^+}\right) \varphi_{n_3}\left(\frac{k_4^+}{k_4^++k_5^+}\right) \\
  &\times m_{n_1}^{\dagger }\left(k_1^++k_2^+\right)m_{n_2}^{\dagger }\left(k_3^++k_5^+\right)m_{n_3}\left(k_4^++k_5^+\right)\notag \\
  \notag  &+\frac{ \delta \left(k_1^++k_2^++k_3^+-k_4^+\right)}{\left(k_1^++k_3^+\right)^2\sqrt{{\left(k_1^++k_2^+\right) \left(k_3^++k_5^+\right) \left(k_4^++k_5^+\right)}}} \varphi_{n_1}\left(\frac{k_1^+}{k_1^++k_2^+}\right) \varphi_{n_2}\left(\frac{k_5^+}{k_3^++k_5^+}\right) \varphi_{n_3}\left(\frac{k_5^+}{k_4^++k_5^+}\right) \\
  &\times m_{n_1}^{\dagger }\left(k_1^++k_2^+\right)m_{n_2}^{\dagger }\left(k_3^++k_5^+\right)m_{n_3}\left(k_4^++k_5^+\right) \notag \\
  \notag  &+\frac{ \delta \left(-k_1^+-k_2^++k_3^+-k_4^+\right) }{\left(k_3^+-k_1^+\right)^2\sqrt{{\left(k_1^++k_2^+\right) \left(k_3^++k_5^+\right) \left(k_4^++k_5^+\right)}}} \varphi_{n_1}\left(\frac{k_2^+}{k_1^++k_2^+}\right) \varphi_{n_2}\left(\frac{k_5^+}{k_3^++k_5^+}\right) \varphi_{n_3}\left(\frac{k_5^+}{k_4^++k_5^+}\right)  \\
  &\times m_{n_1}\left(k_1^++k_2^+\right)m_{n_2}^{\dagger }\left(k_3^++k_5^+\right)m_{n_3}\left(k_4^++k_5^+\right)+\mathrm{h.c.}\Bigg]+\cdots.
\label{eq:V_lf_full}
\end{align}
The ellipsis represents the operators containing three $m^\dagger$ or $m$ and those suppressed by a factor of $1/N_c$ or more,
which are irrelevant to the light-cone GPD.

The three-meson vertex function $\varGamma$ in light-front quantization is defined in \eqref{meson:transition:to:two:meson:matrix:element}.
Its explicit expression reads~\cite{Callan:1975ps}:
\begin{align}
 \notag \varGamma_{n_1,n_2,n_3}\left(x,\bar{x}\right) &=
 4\lambda\sqrt{\frac{\pi}{N_c}}
 \int_{0}^{\bar{x}}dy_1\int_0^{x}dy_2 \frac{1}{\left(y_1+y_2\right)^2} \bigg[\varphi_{n_1}\left(\bar{x}
 y_1\right)\varphi_{n_2}\left(\frac{y_2}{x}\right)\varphi_{n_3}\left(1-\frac{y_1}{\bar{x}}\right)
\\
&\qquad -\varphi_{n_1}\left(x+ y_1\right)\varphi_{n_2}\left(1-\frac{y_2}{x}\right)\varphi_{n_3}\left(\frac{y_1}{\bar{x}}\right)\bigg]+\left(x\leftrightarrow \bar{x} \text{ and } n_2\leftrightarrow n_3\right),
\label{eq:3msnvrtx_LCWF}
\end{align}
with $\bar{x}\equiv1-x$.

\section{Interacting Hamiltonian and three-meson vertex in equal-time quantization}
\label{appendix:B}

As indicated in \eqref{Hamiltonian:re:split:three:pieces} in Sec.~\ref{H:approach:ET:quantization},
we split the Hamiltonian $\mathbb{H}$ in three pieces. There is a piece representing the mesonic interacting Hamiltonian, $\V$,
beginning with ${\cal O}(1/\sqrt{N_c})$.
We have listed one typical term in \eqref{eq:V_incomplete}.
Here we present the complete expression of the $\V$:
 \begin{align}
  \notag \V=\;&\frac{\lambda}{\sqrt{N_c}}\sum_{n_1,n_2,n_3}\int \frac{dk_1dk_2dk_3dk_4dk_5}{(2\pi)^3}\\
  \Bigg[&-\cos \frac{\theta \left(k_1\right)-\theta \left(k_4\right)}{2}\sin \frac{\theta \left(k_2\right)-\theta \left(k_3\right)}{2}m_{n_1}^{\dagger }\left(k_1-k_2\right)m_{n_2}^{\dagger }\left(k_3-k_5\right)m_{n_3}\left(k_4-k_5\right)\frac{1}{\left(k_2-k_3\right)^2}\notag \\
    &\times \delta \left(-k_1+k_2-k_3+k_4\right)\varphi_{n_1}^-\left(k_1,k_1-k_2\right)\varphi_{n_2}^-\left(k_3,k_3-k_5\right)\varphi_{n_3}^-\left(k_4,k_4-k_5\right)\notag \\
    &-\cos \frac{\theta \left(k_2\right)-\theta \left(k_4\right)}{2}\sin \frac{\theta \left(k_1\right)-\theta \left(k_3\right)}{2}m_{n_1}^{\dagger }\left(k_1-k_2\right)m_{n_2}^{\dagger }\left(k_5-k_3\right)m_{n_3}\left(k_5-k_4\right)\frac{1}{\left(k_2-k_4\right)^2}\notag \\
    &\times \delta \left(-k_1+k_2+k_3-k_4\right)\varphi_{n_1}^-\left(k_1,k_1-k_2\right)\varphi_{n_2}^-\left(k_5,k_5-k_3\right)\varphi_{n_3}^-\left(k_5,k_5-k_4\right)\notag \\
    &+\cos \frac{\theta \left(k_1\right)-\theta \left(k_2\right)}{2}\sin \frac{\theta \left(k_3\right)-\theta \left(k_4\right)}{2}m_{n_1}^{\dagger }\left(k_1-k_5\right)m_{n_2}^{\dagger }\left(k_3-k_2\right)m_{n_3}\left(k_4-k_5\right)\frac{1}{\left(k_4-k_3\right)^2}\notag \\
    &\times \delta \left(-k_1+k_2-k_3+k_4\right)\varphi_{n_1}^-\left(k_1,k_1-k_5\right)\varphi_{n_2}^+\left(k_3,k_3-k_2\right)\varphi_{n_3}^-\left(k_4,k_4-k_5\right)\notag \\
    &+\cos \frac{\theta \left(k_1\right)-\theta \left(k_3\right)}{2}\sin \frac{\theta \left(k_2\right)-\theta \left(k_4\right)}{2}m_{n_1}^{\dagger }\left(k_5-k_1\right)m_{n_2}^{\dagger }\left(k_3-k_2\right)m_{n_3}\left(k_5-k_4\right)\frac{1}{\left(k_1-k_3\right)^2}\notag \\
    &\times \delta \left(k_1+k_2-k_3-k_4\right)\varphi_{n_1}^-\left(k_5,k_5-k_1\right)\varphi_{n_2}^+\left(k_3,k_3-k_2\right)\varphi_{n_3}^-\left(k_5,k_5-k_4\right)\notag \\
    &-\cos \frac{\theta \left(k_2\right)-\theta \left(k_3\right)}{2}\sin \frac{\theta \left(k_1\right)-\theta \left(k_4\right)}{2}m_{n_1}^{\dagger }\left(k_5-k_1\right)m_{n_2}^{\dagger }\left(k_2-k_5\right)m_{n_3}\left(k_3-k_4\right)\frac{1}{\left(k_1-k_4\right)^2}\notag \\
    &\times \delta \left(k_1-k_2+k_3-k_4\right)\varphi_{n_1}^+\left(k_5,k_5-k_1\right)\varphi_{n_2}^-\left(k_2,k_2-k_5\right)\varphi_{n_3}^-\left(k_3,k_3-k_4\right)\notag \\
    &-\cos \frac{\theta \left(k_2\right)-\theta \left(k_4\right)}{2}\sin \frac{\theta \left(k_1\right)-\theta \left(k_3\right)}{2}m_{n_1}^{\dagger }\left(k_1-k_5\right)m_{n_2}^{\dagger }\left(k_5-k_2\right)m_{n_3}\left(k_3-k_4\right)\frac{1}{\left(k_2-k_4\right)^2}\notag \\
    &\times \delta \left(-k_1+k_2+k_3-k_4\right)\varphi_{n_1}^+\left(k_1,k_1-k_5\right)\varphi_{n_2}^-\left(k_5,k_5-k_2\right)\varphi_{n_3}^-\left(k_3,k_3-k_4\right)\notag \\
    &-\cos \frac{\theta \left(k_2\right)-\theta \left(k_3\right)}{2}\sin \frac{\theta \left(k_1\right)-\theta \left(k_4\right)}{2}m_{n_1}^{\dagger }\left(k_1-k_2\right)m_{n_2}^{\dagger }\left(k_3-k_5\right)m_{n_3}\left(k_4-k_5\right)\frac{1}{\left(k_2-k_3\right)^2}\notag \\
    &\times \delta \left(-k_1+k_2-k_3+k_4\right)\varphi_{n_1}^-\left(k_1,k_1-k_2\right)\varphi_{n_2}^+\left(k_3,k_3-k_5\right)\varphi_{n_3}^+\left(k_4,k_4-k_5\right)\notag \\
    &-\cos \frac{\theta \left(k_1\right)-\theta \left(k_3\right)}{2}\sin \frac{\theta \left(k_2\right)-\theta \left(k_4\right)}{2}m_{n_1}^{\dagger }\left(k_1-k_2\right)m_{n_2}^{\dagger }\left(k_5-k_3\right)m_{n_3}\left(k_5-k_4\right)\frac{1}{\left(k_2-k_4\right)^2}\notag \\
    &\times \delta \left(-k_1+k_2+k_3-k_4\right)\varphi_{n_1}^-\left(k_1,k_1-k_2\right)\varphi_{n_2}^+\left(k_5,k_5-k_3\right)\varphi_{n_3}^+\left(k_5,k_5-k_4\right)\notag \\
    &+\cos \frac{\theta \left(k_1\right)-\theta \left(k_3\right)}{2}\sin \frac{\theta \left(k_2\right)-\theta \left(k_4\right)}{2}m_{n_1}^{\dagger }\left(k_5-k_1\right)m_{n_2}^{\dagger }\left(k_2-k_5\right)m_{n_3}\left(k_4-k_3\right)\frac{1}{\left(k_4-k_2\right)^2}\notag \\
    &\times \delta \left(k_1-k_2-k_3+k_4\right)\varphi_{n_1}^+\left(k_5,k_5-k_1\right)\varphi_{n_2}^-\left(k_2,k_2-k_5\right)\varphi_{n_3}^+\left(k_4,k_4-k_3\right)\notag \\
    &+\cos \frac{\theta \left(k_1\right)-\theta \left(k_4\right)}{2}\sin \frac{\theta \left(k_2\right)-\theta \left(k_3\right)}{2}m_{n_1}^{\dagger }\left(k_1-k_5\right)m_{n_2}^{\dagger }\left(k_5-k_2\right)m_{n_3}\left(k_4-k_3\right)\frac{1}{\left(k_4-k_1\right)^2}\notag \\
    &\times \delta \left(-k_1+k_2-k_3+k_4\right)\varphi_{n_1}^+\left(k_1,k_1-k_5\right)\varphi_{n_2}^-\left(k_5,k_5-k_2\right)\varphi_{n_3}^+\left(k_4,k_4-k_3\right)\notag \\
    &-\cos \frac{\theta \left(k_3\right)-\theta \left(k_4\right)}{2}\sin \frac{\theta \left(k_1\right)-\theta \left(k_2\right)}{2}m_{n_1}^{\dagger }\left(k_1-k_5\right)m_{n_2}^{\dagger }\left(k_3-k_2\right)m_{n_3}\left(k_4-k_5\right)\frac{1}{\left(k_4-k_3\right)^2}\notag \\
    &\times \delta \left(-k_1+k_2-k_3+k_4\right)\varphi_{n_1}^+\left(k_1,k_1-k_5\right)\varphi_{n_2}^+\left(k_3,k_3-k_2\right)\varphi_{n_3}^+\left(k_4,k_4-k_5\right)\notag \\
    &-\cos \frac{\theta \left(k_2\right)-\theta \left(k_4\right)}{2}\sin \frac{\theta \left(k_1\right)-\theta \left(k_3\right)}{2}m_{n_1}^{\dagger }\left(k_5-k_1\right)m_{n_2}^{\dagger }\left(k_3-k_2\right)m_{n_3}\left(k_5-k_4\right)\frac{1}{\left(k_1-k_3\right)^2}\notag \\
    &\times \delta \left(k_1+k_2-k_3-k_4\right)\varphi_{n_1}^+\left(k_5,k_5-k_1\right)\varphi_{n_2}^+\left(k_3,k_3-k_2\right)\varphi_{n_3}^+\left(k_5,k_5-k_4\right)+\mathrm{h.c.}\Bigg]+\cdots.
\end{align}
The ellipsis represents the operators containing three $m^\dagger$ or $m$ and those suppressed by a factor of $1/N_c$ or more,
which are irrelevant to the quasi-GPDs.

The three-meson vertex function $\widetilde{\varGamma}$ in equal-time quantization is introduced in \eqref{meson:transition:to:two:meson:matrix:element}.
Its special form, when considering a parent meson decaying to two mesons in the parent meson's rest frame,
can be found in Ref.~\cite{Kalashnikova:2001df}. Here, we present the general expression of $\widetilde{\varGamma}$ in terms of the chiral angle and BGWFs:
\begin{align}
  \notag \widetilde{\varGamma}_{n_1,n_2,n_3}(p;P)=&
  \lambda {\left({2 \sqrt{P^2\!+\!\mu^2_{n_1}} \sqrt{p^2\!+\!\mu^2_{n_2}} \sqrt{(P\!-\!p)^2\!+\!\mu^2_{n_3}}\over N_c}\right)^{1/2}}
  \;\dashint \frac{dl_1dl_2}{2\pi}\:{1\over (l_1\!-\!l_2)^2}\nn\\
  \notag \times &\Bigg[\;\sin \frac{\theta\!\left(l_1\right)\!-\!\theta\!\left(l_2\right)}{2} \cos \frac{\theta\!\left(l_1\!\!+\!p\right)\!-\!\theta\!\left(l_2\!\!+\!p\right)}{2} \varphi_{n_2}^-\left(l_2\!\!+\!p,p\right) \varphi_{n_1}^-\left(l_1\!\!+\!p,P\right) \varphi_{n_3}^-\left(l_1,P\!\!-\!p\right)\\
   \notag  &+\cos \frac{\theta \left(l_1\right)-\theta \left(l_2\right)}{2} \sin \frac{\theta \left(l_1+p\right)-\theta \left(l_2+p\right)}{2} \varphi_{n_1}^-\left(l_1+P,P\right) \varphi_{n_2}^-\left(l_2+p,p\right) \varphi_{n_3}^-\left(l_1+P,P-p\right)\\
   \notag  &+\sin \frac{\theta \left(l_1\right)-\theta \left(l_2\right)}{2} \cos \frac{\theta \left(l_1-p\right)-\theta \left(l_2-p\right)}{2} \varphi_{n_1}^-\left(l_2,P\right) \varphi_{n_2}^+\left(l_1,p\right) \varphi_{n_3}^-\left(l_2-p,P-p\right)\\
   \notag  &+\cos \frac{\theta \left(l_1\right)-\theta \left(l_2\right)}{2} \sin \frac{\theta \left(l_1-p\right)-\theta \left(l_2-p\right)}{2} \varphi_{n_2}^+\left(l_1,p\right) \varphi_{n_1}^-\left(l_2-p+P,P\right) \varphi_{n_3}^-\left(l_2-p+P,P-p\right)\\
   \notag  &+\cos \frac{\theta \left(l_1\right)-\theta \left(l_2\right)}{2} \sin \frac{\theta \left(l_1-P\right)-\theta \left(l_2-P\right)}{2} \varphi_{n_1}^+\left(l_2,P\right) \varphi_{n_2}^+\left(l_1,p\right) \varphi_{n_3}^-\left(l_1-p,P-p\right)\\
   \notag  &+\sin \frac{\theta \left(l_1\right)-\theta \left(l_2\right)}{2} \cos \frac{\theta \left(l_1+P\right)-\theta \left(l_2+P\right)}{2} \varphi_{n_1}^-\left(l_1+P,P\right) \varphi_{n_2}^+\left(l_2+p,p\right) \varphi_{n_3}^-\left(l_2+P,P-p\right)\\
   \notag  &+\cos \frac{\theta \left(l_1\right)-\theta \left(l_2\right)}{2} \sin \frac{\theta \left(l_1+p\right)-\theta \left(l_2+p\right)}{2} \varphi_{n_2}^-\left(l_2+p,p\right) \varphi_{n_1}^+\left(l_1+p,P\right) \varphi_{n_3}^+\left(l_1,P-p\right)\\
   \notag  &+\sin \frac{\theta \left(l_1\right)-\theta \left(l_2\right)}{2} \cos \frac{\theta \left(l_1-P\right)-\theta \left(l_2-P\right)}{2} \varphi_{n_1}^+\left(l_2,P\right) \varphi_{n_2}^-\left(l_1,p\right) \varphi_{n_3}^+\left(l_1-p,P-p\right)\\
   \notag  &+\sin \frac{\theta \left(l_1\right)-\theta \left(l_2\right)}{2} \cos \frac{\theta \left(l_1+p\right)-\theta \left(l_2+p\right)}{2} \varphi_{n_1}^+\left(l_1+P,P\right) \varphi_{n_2}^-\left(l_2+p,p\right) \varphi_{n_3}^+\left(l_1+P,P-p\right)\\
   \notag  &+\cos \frac{\theta \left(l_1\right)-\theta \left(l_2\right)}{2} \sin \frac{\theta \left(l_1+P\right)-\theta \left(l_2+P\right)}{2} \varphi_{n_1}^-\left(l_1+P,P\right) \varphi_{n_2}^-\left(l_2+p,p\right) \varphi_{n_3}^+\left(l_2+P,P-p\right)\\
   \notag  &+\sin \frac{\theta \left(l_1\right)-\theta \left(l_2\right)}{2} \cos \frac{\theta \left(l_1-p\right)-\theta \left(l_2-p\right)}{2} \varphi_{n_2}^+\left(l_1,p\right) \varphi_{n_1}^+\left(l_2-p+P,P\right) \varphi_{n_3}^+\left(l_2-p+P,P-p\right)\\
   \notag  &+\cos \frac{\theta \left(l_1\right)-\theta \left(l_2\right)}{2} \sin \frac{\theta \left(l_1-p\right)-\theta \left(l_2-p\right)}{2} \varphi_{n_1}^+\left(l_2,P\right) \varphi_{n_2}^+\left(l_1,p\right) \varphi_{n_3}^+\left(l_2-p,P-p\right)\\
   &+\left(p\leftrightarrow P-p \text{ and } n_2\leftrightarrow n_3\right)\Bigg].
\label{eq:3msnvrtx_BGWF}
\end{align}

In the $P\to \infty $ limit,  the backward-moving BGWF $\varphi_n^-(k,P)$ vanishes, and the forward-moving BGWG approaches the LCWF,
$\varphi_n^+(k,P) \to \sqrt{2\pi\over P} \varphi_n\left(\frac{k}{P}\right)$, as indicated in \eqref{eq:BGasy2}.
Therefore only the last three lines in \eqref{eq:3msnvrtx_BGWF},
which entail the product of three $\varphi^+$ wave functions, survive in the infinite boost limit.
Rewriting the momenta as  $l_1\!=\!y_2 P$, $l_2\!=\!-y_1 P$, $p\!=\!x P$ and $l_1\!=\!(x\!-\!y_2)P$, $l_2\!=\!(x+y_1)P$, and
exploiting the limiting forms of the trigonometric functions from \eqref{eq:BGasy}
\begin{align}
\sin\frac{\theta(l_1)-\theta(l_2)}{2} \rightarrow \frac{\epsilon(y_1)+\epsilon(y_2)}{2},\quad\cos\frac{\theta(l_1-p)-\theta(l_2-p)}{2} \rightarrow \frac{\epsilon(x+y_1)\epsilon(x-y_2)+1}{2},
\end{align}
one immediately verifies that the three-meson vertex function $\widetilde{\varGamma}$ in
\eqref{eq:3msnvrtx_BGWF} reduces to its IMF counterpart, \eqref{eq:3msnvrtx_LCWF}.

\section{Comparison between quasi-PDF in this work and in Ref.~\cite{Jia:2018qee}}
\label{appendix:C}

\begin{figure}[htb]
   \centering
   \includegraphics[width=0.875\linewidth]{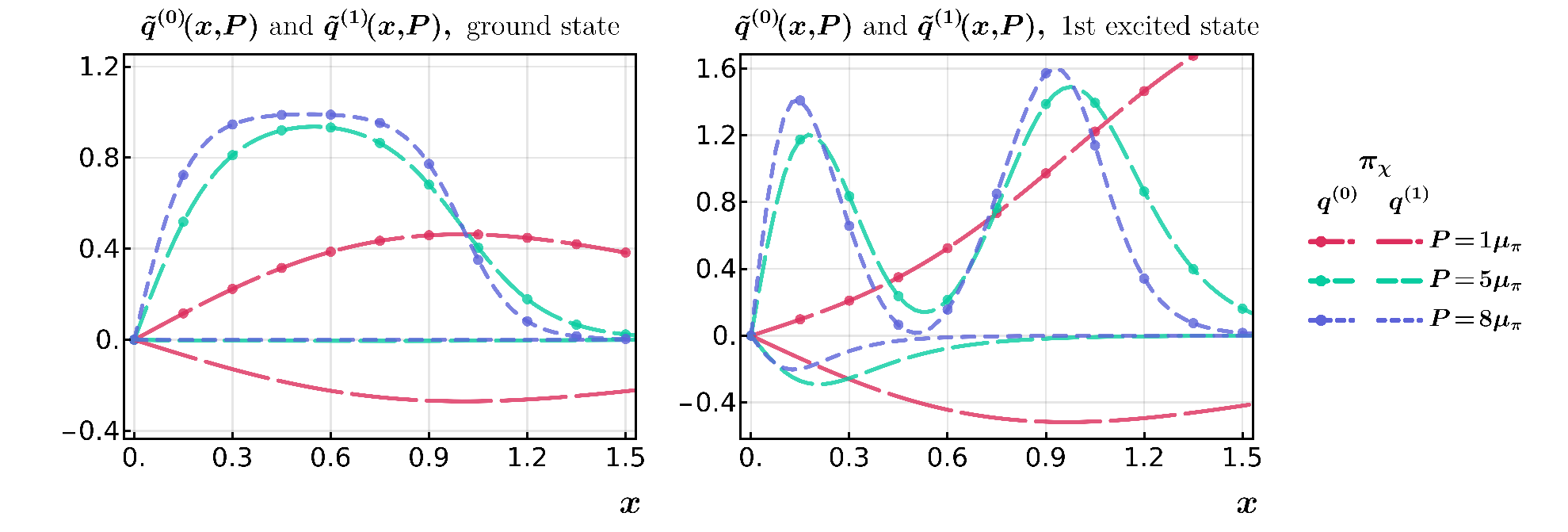}
   \includegraphics[width=0.875\linewidth]{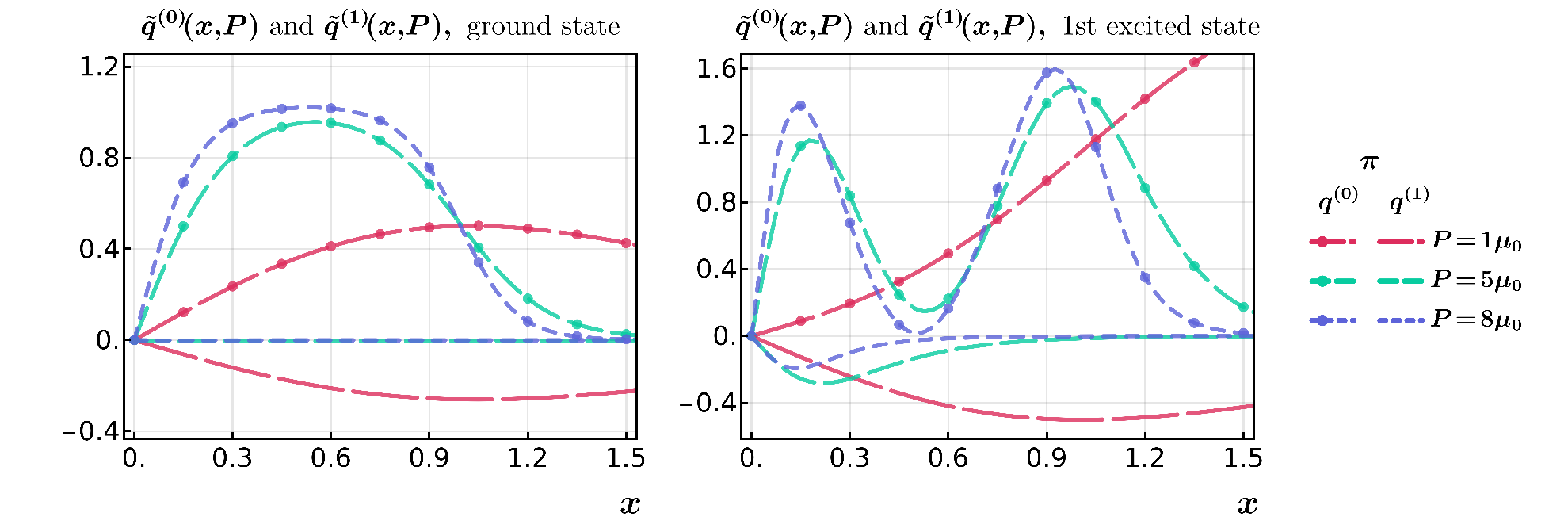}
   \caption{Comparison between $\tilde{q}^{(1)}(x,P)$ and $\tilde{q}^{(0)}(x,P)$, for the light mesons ($\pi_\chi$ and $\pi$)}\label{fig:q0vsq1_A}
   \end{figure}

\begin{figure}[htb]
     \centering
     \includegraphics[width=0.875\linewidth]{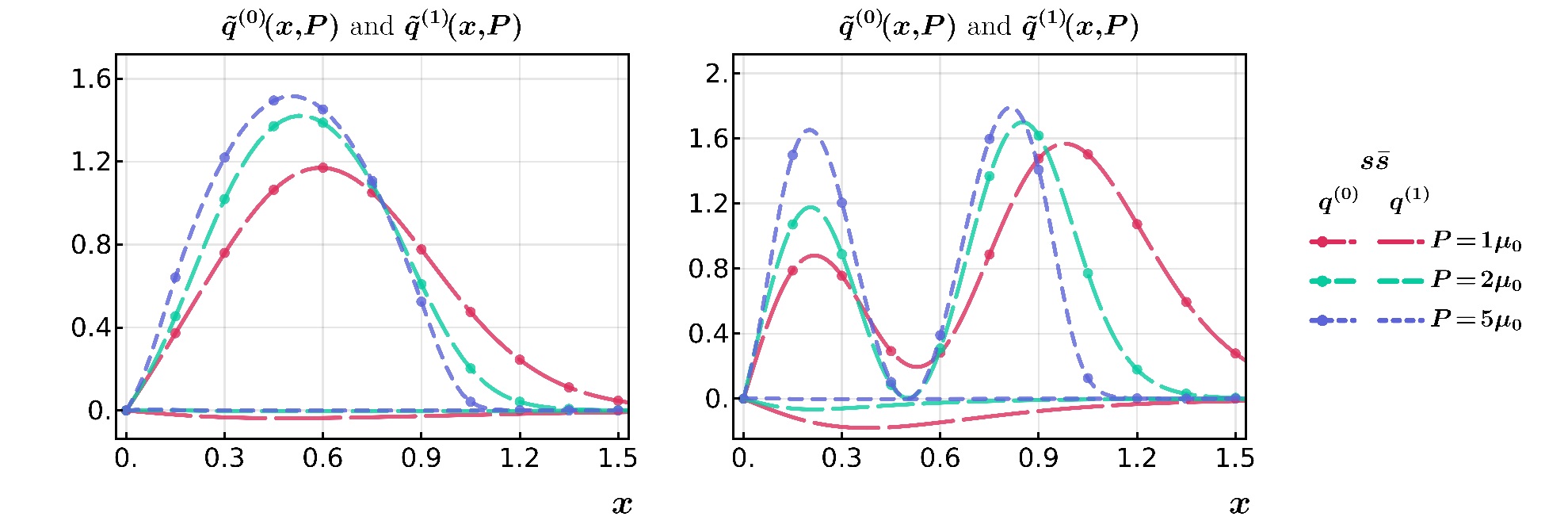}
     \includegraphics[width=0.875\linewidth]{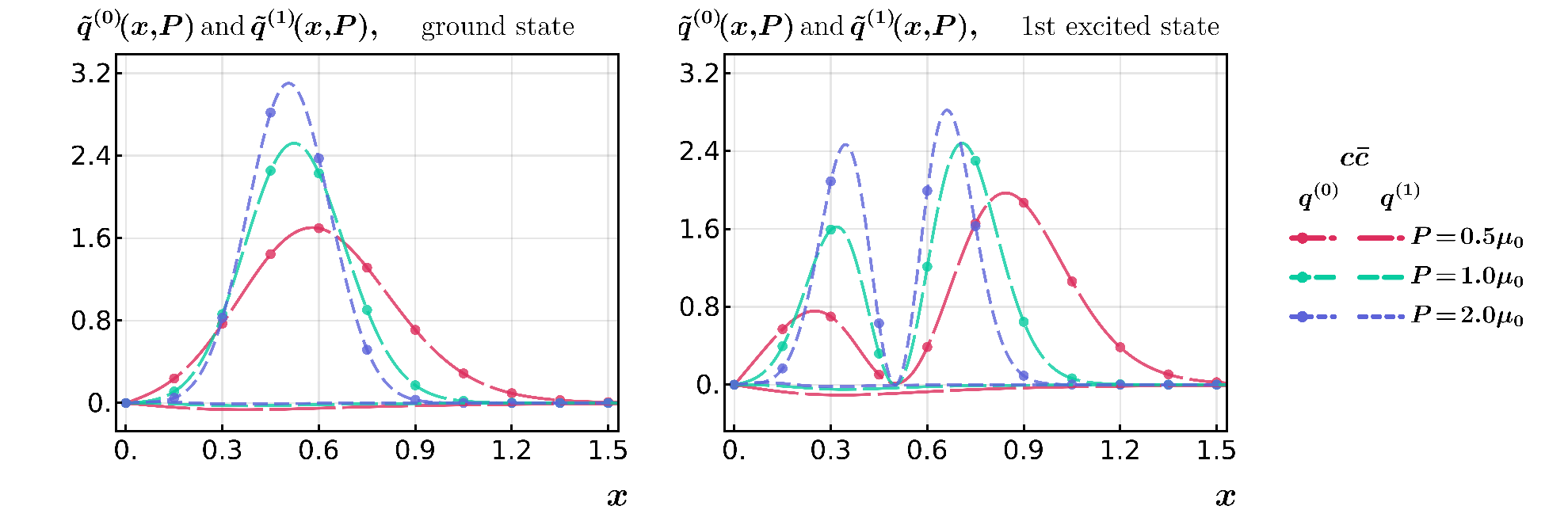}
     \caption{Same as Fig.~\ref{fig:q0vsq1_A}, but the mesons are replaced with the $s\bar{s}$ and $c\bar{c}$ quarkonia.}
     \label{fig:q0vsq1_B}
     \end{figure}

The complete ${\cal O}(N_c^0)$ contribution of the quasi quark PDF of a flavor-neutral meson is decomposed into
$\tilde{q}_n(x,P) = \tilde{q}_n^{(0)}(x,P)+\tilde{q}_n^{(1)}(x,P)$, as indicated in \eqref{quasi:PDF:break:into:q0:q1}.
Here we make a numerical assessment on the importance of the new $\tilde{q}_n^{(1)}(x,P)$ piece
with respect to the old incomplete expression for quasi-PDF, $\tilde{q}_n^{(0)}(x,P)$~\cite{Jia:2018qee}.
In Figs.~\ref{fig:q0vsq1_A} and~\ref{fig:q0vsq1_B} we juxtapose $\tilde{q}^{(1)}(x,P)$ and $\tilde{q}^{(0)}(x,P)$ in each plot for different types of mesons.
One can clearly observes that, for light meson cases ($\pi_\chi$ and $\pi$), $\tilde{q}^{(1)}(x,P)$ is comparable in magnitude with $\tilde{q}^{(0)}(x,P)$
when the light meson carries soft momentum.
With the increasing quark mass, the effect of $\tilde{q}^{(1)}(x,P)$ becomes negligible compared to $\tilde{q}^{(0)}(x,P)$,
which is exemplified by the $s\bar{s}$ and $c\bar{c}$ mesons
In addition, the $q^{(1)}(x,P)$ appears to be more sizable for the first excited-state meson than that for the ground state.
In all cases, $\tilde{q}^{(1)}(x,P)$ quickly fade away when the meson is boosted to large momentum.


\end{document}